\documentclass[11pt]{article}
\usepackage{graphics}
\usepackage{epsfig}

\newcommand{\be}{\begin{eqnarray}}
\newcommand{\ee}{\end{eqnarray}}

\def\ll#1{\left#1}
\def\r#1{\right#1}
\def\fr{\frac{1}{2}}

\def\mref#1{(\ref{#1})}

\def\bd{\begin{displaymath}}
\def\ed{\end{displaymath}}

\def\ba#1{\begin{array}{#1}}
\def\ea{\end{array}}
\def\nn{\nonumber}
\newfont{\Bbb}{msbm10 scaled 1200}

\begin{document}

\pagestyle{empty}

\begin{center}

{\LARGE\bf Superscar states in rational polygon billiards - a reality or an illusion?\\[0.5cm]}

\vskip 12pt

{\large {\bf Stefan Giller}}

\vskip 3pt

Jan D{\l}ugosz University in Czestochowa\\
Institute of Physics\\
Armii Krajowej 13/15, 42-200 Czestochowa, Poland\\
e-mail: stefan.giller@ajd.czest.pl
\end{center}

\vspace{30pt}

$\;\;\;\;\;\;\;\;\;\;\;\;\;\;\;\;\;\;\;\;\;\;\;\;\;\;\;\;\;\;\;\;\;\;\;\;\;\;\;\;\;\;\;\;\;\;\;\;\;\;\;\;\;\;\;\;\;\;\;\;\;\;\;\;\;\; Moim\;wnukom,\;Guciowi\;i\;Tytuskowi,$

$\;\;\;\;\;\;\;\;\;\;\;\;\;\;\;\;\;\;\;\;\;\;\;\;\;\;\;\;\;\;\;\;\;\;\;\;\;\;\;\;\;\;\;\;\;\;\;\;\;\;\;\;\;\;\;\;\;\;\;\;\;\;
\;\;\;\;\;\;\;\;\;\;\;\;\;\;\;\;\;\;\;\; lubi${\it\c{a}}$cym\;rysunki$
\vspace{20pt}

\begin{abstract}
The superscars phenomena (Heller, E.J., {\it Phys. Rev. Lett.} {\bf 53}, (1984) 1515) in the rational polygon billiards (RPB) are analysed using the high
energy semiclassical wave functions (SWF) built on classical trajectories forming skeletons. Considering examples of the pseudointegrable billiards such
as the Bogomolny-Schmit triangle, the parallelogram and the L-shape billiards as well as the integrable rectangular one the constructed SWFs allow us to
verify the idea of Bogomolny and Schmit ({\it Phys. Rev. Lett.} {\bf 92} (2004) 244102) of SWFs (superscars) propagating along periodic orbit channels
(POC) and vanishing outside of them. It is shown that the superscars effects in RPB appear as natural properties of SWFs built on the periodic skeletons.
The latter skeletons are commonly present in RPB and are always composed of POCs. The SWFs built on the periodic skeletons satisfy all the basic
principles of the quantum mechanics contrary to the superscar states of Bogomolny and Schmit which break them. Therefore the superscars effects need not
to invoke the idea of the superscar states of Bogomolny and Schmit at least in the cases considered in our paper.
\end{abstract}

\vskip 3pt
\begin{tabular}{l}
{\small PACS number(s): 03.65.-w, 03.65.Sq, 02.30.Jr, 02.30.Lt, 02.30.Mv} \\[1mm]
{\small trajectories, chaotic dynamics, quantum chaos, scars, superscars}
\end{tabular}

\newpage

\pagestyle{plain}

\setcounter{page}{1}

\section{Introduction}

\hskip+2em Bogomolny and Schmit have invented the idea \cite{46} that POCs are the special periodic skeletons in RPB in which some untypical (deprived of
a smoothness in RPB) semiclassical solutions to
the Schr{\"o}dinger equation can exist and propagate along them and which can resonate and accompany the typical ones, i.e. these which satisfy the basic
conditions for the wave
functions inside billiards such as the continuity and the smoothness and some well defined boundary conditions. The authors came to their idea extrapolated the results of their calculations on the semiclassical
limit of diffraction of the electromagnetic waves
on a staggered periodic array of the conducted half-planes to the same semiclassical limit of the polygon billiards quantum mechanics \cite{45}. According to the authors
POCs can carry
SWFs inside them called superscars \cite{44} satisfying on the POCs boundaries the Dirichlet conditions. They supported
their conjecture by respective numerical calculations for particular energies just using to this goal the example of the pseudointegrable right triangle
shown in Fig.1 of the present paper. In
particular they argued that forms of the exact wave functions they calculated numerically for the triangular billiards of Fig.1 show clearly the structure
corresponding to excitations of the wave functions of the POC with the period ${\bf D}_6$. In particular a position of the exact wave function nodal
lines was to support the claims of the authors.

In this paper we are trying to verify the Bogomolny - Schmit conjecture by constructing explicitly SWFs for several RPB and analyzing a dependence of
their internal structures on the one of the corresponding elementary polygon pattern (EPP) the latter being a basic brick of the rational billiard
Riemann surface (RBRS) on which SWFs are defined \cite{43}-\cite{42}. In particular SWFs are built on periodic skeletons of respective RBRS which always are composed of POCs.
We consider in this way SWFs built in the right triangle investigated by Bogomolny and Schmit, in the parallelogram billiards and in the rectangular and
L-shape ones.

The SWFs we have used are constructed on the skeletons which the method is just the RPB version of the Maslov - Fedoriuk approach to the semiclassical
limit
in the quantum mechanics \cite{4}. The explicit application of the method to RPB show that it has some limitations due to the following two basic facts
\begin{enumerate}
\item RPB are pseudointegrable plane periodic systems with a large number ($>2$) of independent periods making impossible their standard semiclassical
quantization - only some special cases of these systems called doubly rational polygon billiards (DRPB) allow for such a quantization and only for a
limited set of their possible states; and
\item the high energy SWFs which can be built in the case of DRPB can be formed by the plane waves only which the circumstance is a further limitation
in constructing a set of the semiclassical solutions to the Schr{\"o}dinger equation satisfying some definite boundary conditions.
\end{enumerate}

Nevertheless the solutions which are provided by the method seem to be
sufficient to get some general conclusions about their internal structure particularly if they are built on the periodic skeletons.

It is just the analysis of the SWFs built on periodic skeletons which show that the superscars effects in the considered RPB including the particular
case of the Bogomolny - Schmit triangle are results of the definite structure of SWFs built on such skeletons. Namely, the periodic skeletons
which are common in the pseudointegrable billiards are always built of some number of POCs. The SWFs calculated on these skeletons and satisfying all
necessary conditions of the solution to the Schr{\"o}dinger equation (continuity, smoothness and boundary conditions) get then coherent contributions from
all their component POCs which the contributions mimic then the behaviour of the Bogomolny-Schmit states for some sets of energies from the energy spectrum
of the billiards. Each such a contribution alone however does not satisfy the necessary conditions mentioned to be a proper solution to the Schr{\"o}dinger
equation and therefore cannot exist as a quantum state in the billiards considered.

The paper is organized as follows.

In the next section the method of construction of SWFs used in the paper is briefly recognized in the case of the DRPB.

In sec.3 a necessary extension of the method to quantize in arbitrary polygon billiards is discussed.

In sec.4 the semiclassical quantizations in the triangular, parallelogram, rectangular and L-shape billiards are performed and
analysed.

In the case of the triangular billiards the numerical results provided by Bogomolny and Schmit \cite{46} are used to show the numerical
accuracy of the SWFs built for this case.

The case of the rectangular billiards provides us with the rare example of the pure Bogomolny-Schmit superscar
states realized all as so called bouncing ball states which however do satisfy all the quantum mechanical rules. The equilateral triangle is another
such a case not considered however in this paper.

In sec.5 we summarize and discuss the results of the paper.

\section{Semiclassical wave functions in the rational polygon billiards - r\'esum\'e}

\hskip+2em Let us begin with making a r\'esum\'e of our results \cite{43,41,42} on the semiclassical quantization in the rational polygon billiards (RPB) by the
method of Maslov {\it et al} \cite{4} in order to stress their most important points.

The method contains two sectors - the classical and the quantum ones which in the case of RPB can be
summarised by the following points.

\subsection{The classical sector}

\begin{enumerate}
\item any motion in RPB, i.e. in a polygon billiards each angle of which is a rational part of $\pi$, can be considered on a kind of a Riemann surface
called rational polygon Riemann surface (RPRS) obtained by an operation called unfolding which means infinitely repeating mirror reflections of the
billiards by any of its sides;
\item any trajectory of the billiards ball which collides elastically with the billiards boundary is transformed into a straight line on RPRS when
unfolded;
\item any RPRS
\begin{enumerate}
\item is periodic - its periods are defined by all periodic orbits of the corresponding RPB in the following way
\begin{enumerate}
\item every periodic orbit generates a number of periods equal to a number of reflections of the orbits off the billiards boundary;
\item a length of each such a period is the same and equal to a length of the corresponding periodic orbit;
\item directions of periods generated by a periodic orbit coincide with the corresponding directions of links of a periodic orbit;
\item all other periods of RPRS are vectors of the vector space spanned by the set of all periods generated by all periodic orbits with integers as
coefficients;
\end{enumerate}
\item coincides with the plane when the classical motions in RPB are integrable;
\item consists of infinitely many branches in the case of RPBs with classically non-integrable motions but covered in such cases by periodically repeated
finite system of RPBs called elementary polygon pattern (EPP);
\end{enumerate}
\item a set of all periods of RPRS is uniquely defined by the respective RPB;
\item each periodic orbit on RPRS is parallel to some straight line called singular diagonal (SD) which crosses at least two vertices on the
corresponding RPRS;
\item all trajectories on RPRS parallel to each other but none of which crosses any vertex of the polygon billiards considered form a set called a
skeleton;
\item a skeleton which straight line trajectories are parallel to some period are called periodic if all their trajectories are periodic with the same
period and the skeleton itself is bounded by some two singular diagonals - such a periodic skeleton is known as POC \cite{46};
\item an aperiodic skeleton on RPRS is built of trajectories which form subsets of them such that each trajectory of a subset goes by the same set of
points of the billiards and each such a subset itself is dense in the skeleton;
\item a straight line parallel to an aperiodic skeleton but crossing some single vertex on the RPRS does not enter the skeleton (see p. 6) and a full set
of such singular lines is dense on the RPRS - the skeleton and the corresponding set of singular lines exhaust the set of all straight lines of
a given direction on the corresponding RPRS;
\item each RPB with rational angles $\alpha_k=\frac{p_k}{q_k}\pi,\;k=1,...,n$, where $p_k$ and $q_k$ are coprime integers is classically pseudointegrable,
i.e. a classical motion in it realizes in the phase space by trajectories lying on a surface topologically equivalent to a multi-torus with a genus $g$
given by \cite{53}
\be
g=1+\frac{C}{2}\sum_{k=1}^n\frac{p_k-1}{q_k}
\label{1}
\ee
where $n$ is a number of the polygon vertices and $C$ is the least common multiple of all $q_k$, $k=1,...,n$;
\item for rare cases when $p_k=1$ for each $k$ the respective polygons billiards (e.g. the equilateral triangle ones, the rectangle ones) are integrable;
\item a vector space of all periods of RPRS is at most the $2g$-dimensional vector space $V_{2g}$ over integers, i.e. there are at most $2g$ linearly
independent periods ${\bf D}_k,\;k=1,...,2g$, such that
\begin{enumerate}
\item none of the periods ${\bf D}_k,\;k=1,...,2g$, can be expressed as a linear combination of the others with integer coefficients; and
\item any other period belonging to $V_{2g}$ is a linear combination of them with integers as coefficients;
\end{enumerate}
\item not every vector of  $V_{2g}$ corresponds to a periodic orbit - most of them are responsible only for the periodic structure of RPRS;
\item a periodic structure of RPRS is realized by a figure called elementary polygon pattern (EPP) which contains $2C$ possible
different positions of the original RPB obtained by its mirror reflections in their sides;
\item on any of EPP realizing RPRS one can find $2g$ independent periods of the space $V_{2g}$ as those which connect pairs of respective edges of EPP
parallel to each other;
\item each RPRS can be projected on a plane together with the vector space $V_{2g}$ of its periods;
\item considered on a real plane a projection of the space $V_{2g}$ becomes two dimensional so that in any set of $2g$ of its independent periods only
two of them, say ${\bf D}_k,\;k=1,2$, can be independent while the remaining $2g-2$ ones, ${\bf D}_{2+k},\;k=1,...,2g-2$, can be expressed by the
former by the following linear combinations
\be
{\bf D}_{2+k}=a_{k1}{\bf D}_1+a_{k2}{\bf D}_2,\;\;\;\;k=1,...,2g-2
\label{1a}
\ee
where the following possibilities for the coefficients $a_{ki}$ can happen
\begin{enumerate}
\item the two independent periods ${\bf D}_k,\;k=1,2$, are such that all $a_{ki}$ are integers;
\item the linear combinations \mref{1a} contains rationals as coefficients independently of the choice of ${\bf D}_k,\;k=1,2$ (the case of so called
doubly rational polygon billiards (DRPB) \cite{42});
\item the linear combinations \mref{1a} contains irrationals as coefficients independently of the choice of ${\bf D}_k,\;k=1,2$;
\end{enumerate}
\item in the case of DRPB for any three of the $2g$ linearly independent periods ${\bf D}_k,\;k=1,...,2g$, of the space $V_{2g}^{DRPB}$ we have
\be
\frac{p_i}{q_i}{\bf D}_i+\frac{p_j}{q_j}{\bf D}_j+\frac{p_k}{q_k}{\bf D}_k=0\;\;\;\;\;\;i,j,k=1,...,2g,\;i\neq j\neq k
\label{1b}
\ee
where $p_i,p_j,p_k$ are coprime pairwise as well as $q_i,q_j,q_k$.
\item if it is possible to choose from the set ${\bf D}_k,\;k=1,...,2g$ such two periods say ${\bf D}_1$ and ${\bf D}_2$, for which
their respective coefficients $q_1,q_2$ in \mref{1b} are always different for any remaining ${\bf D}_j,\;j=3,...,2g$ in the relation \mref{1b} so that
we have
\be
{\bf D}_j=\frac{p_{1j}q_j}{q_{1j}p_j}{\bf D}_1+\frac{p_{2j}q_j}{q_{2j}p_j}{\bf D}_2\;\;\;\;\;\;j=3,...,2g-2
\label{1c}
\ee
then $\frac{{\bf D}_i}{C_i},\;i=1,2$, with $C_i$ being the least common multiple of the respective denominators $q_{ij}p_j,\;i=1,2,\;j=3,...,2g-2$,
in \mref{1c}, are also periods of RPRS, i.e. belong to $V_{2g}^{DRPB}$ (see App.A). Other cases than the considered ones have to be treated separately;
\item if the linear combinations \mref{1a} contains irrationals none two dimensional base exists in the projected $V_{2g}$ - the projection contains
arbitrary small periods so that the corresponding EPPs which the respective RPRS is glued from densely cover the plane;

\subsection{The quantum sector}

\item semiclassical wave functions (SWF) satisfying some boundary conditions in a RPB can be built selfconsistently only in the DRPB cases both on
the periodic and aperiodic skeletons by linear combinations of so called basic semiclassical wave functions (BSWF) defined on the corresponding RPRS in
the following form
\be
\Psi_{BSWF}^\pm(x,y,p_x,p_y)=e^{\pm i\lambda(p_xx+p_yy)}\chi_{BSWF}^\pm(x,y,p)
\label{2}
\ee
where $\lambda=\hbar^{-1}$ (and will be
put further equal to 1 as well as the billiard ball mass), $p$ is a value of the billiard ball classical momentum
${\bf p}=[p_x,p_y]$ and the factors $\chi^\pm(x,y,p)$ are given by the following semiclassical series
for $p\to +\infty$:
\be
\chi_{BSWF}^\pm(x,y,p)=\sum_{k\geq 0}\frac{\chi_k^\pm(x,y)}{p^k}
\label{3}
\ee
while the corresponding semiclassical energy spectrum is searched in the form of the series
\be
E=\fr p^2+\sum_{k\geq 0}\frac{E_k}{p^k}
\label{4}
\ee
\item $\Psi_{BSWF}^\pm(x,y,p_x,p_y)$ must be periodic on the projected RPRS with respect to all its periods which demand can be satisfied exactly only in the
case of DRPB due to the property 18. above;
\item SWFs in DRPB and the quantum states they describe can exist in two forms depending which kind of the skeletons, periodic or aperiodic, they are
defined on
\begin{enumerate}
\item in the aperiodic case of the skeleton its property $8.$ above determines the factors \mref{3} to be constant (put then arbitrarily
equal to $1$) while the series in \mref{4} disappears leaving the energy $E$ to be equal totally to the kinetic energy of the billiards ball, i.e.
\be
\chi^\pm(x,y,p)\equiv 1\nn\\
E=\fr p^2
\label{5}
\ee
\item in the case of the periodic skeleton with its period perpendicular to the $x$-axis the factors $\chi^\pm(x,y,p)$ have the following general form
\be
\chi^\pm(x,y,p)=A\cos(\sqrt{2E_0}x)+B\sin(\sqrt{2E_0}x)
\label{6}
\ee
where $E_0$ is determined by periodicity of $\chi^\pm(x,y,p)$ on RPRS while the corresponding energy $E$ is given by
\be
E=\fr p^2+E_0
\label{7}
\ee
with the condition
\be
\sqrt{2E_0}<<p
\label{7a}
\ee
stressing the asymptotic origin of \mref{7};
\end{enumerate}
\item the semiclassical energy spectrum covered by the method presented is determined totally by its periods ${\bf D}_i,\;i=1,2$, by the following formulas
\be
{\bf p}\cdot{\bf D}_1=2\pi C_1m,\;\;\;\;\;\;\;\;\;\;\;\;\;\;\;\;\;\;\;\;\;\;\;\;\;\;\;\;\;\;\;\;\;\;\;\;\;\nn\\
{\bf p}\cdot{\bf D}_2=2\pi C_2n,\;\;\;\;\;\;\;m,n=0,\pm 1,\pm 2,...
\label{8}
\ee
so that
\be
{\bf p}_{mn}=2\pi\frac{(mC_1{\bf D}_2-nC_2{\bf D}_1)\times({\bf D}_1\times{\bf D}_2)}{({\bf D}_1\times{\bf D}_2)^2}=\nn\\
2\pi\frac{mC_1D_2^2-nC_2{\bf D}_1\cdot{\bf D}_2}{({\bf D}_1\times{\bf D}_2)^2}{\bf D}_1+
2\pi\frac{nC_2D_1^2-mC_1{\bf D}_1\cdot{\bf D}_2}{({\bf D}_1\times{\bf D}_2)^2}{\bf D}_2\nn\\
|m|+|n|>0,\;\;\;\;\;\;\;\;\;\;\;\;m,n=0,\pm 1,\pm 2,...
\label{9}
\ee
and
\be
E_{mn}=\frac{{\bf p}_{mn}^2}{2}=2\pi^2\frac{|mC_1{\bf D}_2-nC_2{\bf D}_1|^2}{|{\bf D}_1\times{\bf D}_2|^2}
,\;\;\;\;\;\;\;m,n=\pm 1,\pm 2,...
\label{10}
\ee
with the following restrictions
\begin{enumerate}
\item both the spectrum and the corresponding SWFs are exact for the case $17(a)$ of the relations between independent periods and are complete for the
integrable RPB and may be incomplete in the opposite case;
\item both the spectrum and the corresponding SWFs are exact but are not complete in the case of DRPB (the case $17(b)$);
\item in the case $17(c)$ the semiclassical energy levels and the corresponding SWFs can be found only approximately (approximating irrationals by
rationals in \mref{1a}) and are not complete - the more accurate are rationals approximating the corresponding irrationals the more rare are levels
provided by the method and the higher regions of the spectrum they occupy;
\end{enumerate}
\item in the periodic skeleton case a possibility to construct on it a respective SWF can depend on a geometry of the corresponding DRPB demanding some relations
to exist between its periods and goes as follows
\begin{enumerate}
\item the kinetic contribution to the energy is given by the condition
\be
{\bf p}\cdot{\bf D}_2=pD_2=2\pi nC_2,\;\;\;\;\;\;\;n=\pm 1,\pm 2,...
\label{11}
\ee
where ${\bf D}_2$ is the period of the skeleton;
\item assuming the $x$-axis to be perpendicular to ${\bf D}_2$, $E_0$ can be defined by the second period ${\bf D}_1/C_1$ by
\be
E_{0m}=\frac{2\pi^2m^2C_1^2}{D_1^2\sin^2\alpha},\;\;\;\;\;\;\;n=0,\pm 1,\pm 2,...
\label{11a}
\ee
where $\alpha$ is an angle made by ${\bf D}_1$ and ${\bf D}_2$ which satisfy the following conditions
\be
\frac{C_2D_1\cos\alpha}{C_1D_2}=\frac{r}{s}
\label{11b}
\ee
for some coprime integers $r,s$ and for each choice of ${\bf D}_1$ which the condition is satisfied for DRPB;
\item the spectrum is then given by
\be
E_{mn}=\fr{\bf p}^2+E_{0m}=2\pi^2\ll(\frac{n^2C_2^2}{D_2^2}+\frac{m^2C_1^2}{D_1^2\sin^2\alpha}\r)\;\;\;\;\;\;\;m,n=0,\pm 1,\pm 2,...
\label{12}
\ee
with the condition
\be
\frac{m}{n}<<\frac{D_1C_2\sin\alpha}{D_2C_1}
\label{12a}
\ee
as the formal condition demanded by the asymptotic form of \mref{12};
\end{enumerate}
\item due to the conditions \mref {8}, \mref{11} and \mref{11a} SWFs \mref{2} are periodic on the whole RPRS corresponding to the considered case of DRPBs;
\item if the quantization on a periodic skeleton is possible then both the energy spectrum and the corresponding SWF constructed for such a case can be
obtain also from a general formulas for the aperiodic skeletons
\item a semiclassical wave function on an aperiodic skeleton corresponding to a given energy of the spectrum and satisfying
some billiards boundary conditions can be built totally by a linear combination of all plane waves given by \mref{2} and corresponding to all momenta
arising from the classical reflections of the billiards ball by the billiards boundary
\be
\Psi^\pm(x,y,p_x,p_y)=\sum_{k=1}^{2C}\eta_k e^{\pm i(p_xx_k+p_yy_k)},\;\;\;\;\eta_k=\pm
\label{13}
\ee
where a set $\eta_k,\;k=1,...,2C$ of signs depend on kinds of boundary conditions allowed by the EPP corresponding to the RPB considered and
$(x_k,y_k),\;k=1,...,2C$, are coordinates of all positions of the original point of the billiards obtained by the mirror reflections of the RPB to get
its EPP;
\item SWFs in the case of periodic skeletons is constructed by the same formula \mref{13} with the condition that the corresponding BSWFs defined on each
of the parallel skeletons both periodic and aperiodic ones running through the considered EPP are properly matched according to the rules of the quantum
mechanics;
\item not all sets of $\eta_k,\;k=1,...,2C$, and consequently not all set of boundary conditions on the RPB sides are allowed
- possible boundary conditions are completely determined by a topology of EPP corresponding to the billiards considered and cannot be arbitrary - a restriction
which is also strictly related to the fact that the corresponding SWFs are built by the plane waves only.
\end{enumerate}

\section{Semiclassical wave functions in an arbitrary polygon billiards}

\hskip+2em SWF cannot be constructed directly in the polygon billiards which are rationals but not doubly rationals as well as in billiards which are
irrationals, i.e. if some angles of the latter are irrational when measured in the $\pi$-unit. In both the cases we have to make additional approximations
which allow us to reduce the cases to the respective DRPBs. Such a reduction appeals to general theorems describing the conditions put on boundaries of
two billiards to make their energy spectra arbitrarily close to each other in a controlled way (see \cite{40} and \cite{42}, App.C).

In the case of RPB which are not DRPB the coefficients in the relations \mref{1b} which are irrational are approximated by respective rationals allowing
to construct on the respective EPP SWFs which satisfy the allowed boundary conditions only approximately but with a desired accuracy. Denote this
reduction by RPB$\to$DRPB.

In the case of irrational polygon billiards (IPB) the following two steps in the respective approximation are necessary
\begin{enumerate}
\item the one which substitutes an IRB by a respective RPB with a desired accuracy (denote it by IPB$\to$RPB); and
\item the second which makes from RPB obtained in the previous step a corresponding DRPB in the way describe just above (RPB$\to$DRPB).
\end{enumerate}

In the next section the approximate procedure mentioned above will be applied to several examples of RPB to build on it the semiclassical wave
functions satisfying allowed boundary conditions and to get the corresponding energy spectra.

\section{Do superscar phenomena in RPB need superscar states developed in POCs?}

\hskip+2em In this section we consider several examples of RPBs to show that the superscar phenomena observed in the billiards considered, i.e. in the
triangle of the Bogomolny and Schmit \cite{46}, in the parallelogram billiards and in the triangular one and the L-shape billiards do not need to
invoke effects of SWFs defined in POCs as it was suggested by Bogomolny and Schmit. In particular it will be clarified why in the cases of the rectangular billiards
where POCs are abundant and the L-shape billiards the only visible superscars corresponds to the bouncing ball skeletons.

\subsection{The Bogomolny - Schmit triangle}

\hskip+2em The triangle is right with the remaining angle equal to $\pi/8$ and $3\pi/8$. A motion in it is pseudointegrable on a two holes torus, i.e.
with $g=2$.
Its corresponding EPP is shown in Fig.1 together with its three periodic copies. The relations between the periods of the respective RBRS shown in Fig.1
prove that the billiards is not DRPB and to quantize it semiclassically we need to approximate irrational coefficients in these relations by some
rational ones close to them. As it was discussed by Gutzwiller \cite{3} such an approximation is not a trivial problem which have its own limitations
depending on a type of an irrational, i.e. whether it is algebraic or transcendental. It follows from Fig.1 that it is enough to approximate $\sqrt{2}$
by some rational $u/q$. It is however important that an accuracy $\epsilon$ of such an approximation defined by $|\sqrt{2}-u/q|<\epsilon$ should be
proportional to some power of $1/q$ clearly smaller than one. For example if $u/q$ is got by cutting the continued fraction corresponding to $\sqrt{2}$
then $\epsilon=1/(2q^2)$($=1/u^2$ up to $\epsilon^2$). In fact this approximation for $\sqrt{2}$ can be considered as the best one since in this case $\epsilon>1/(3\sqrt{2}q^2)$ for
any $q$ \cite{3}.

\begin{figure}
\begin{center}
\psfig{figure=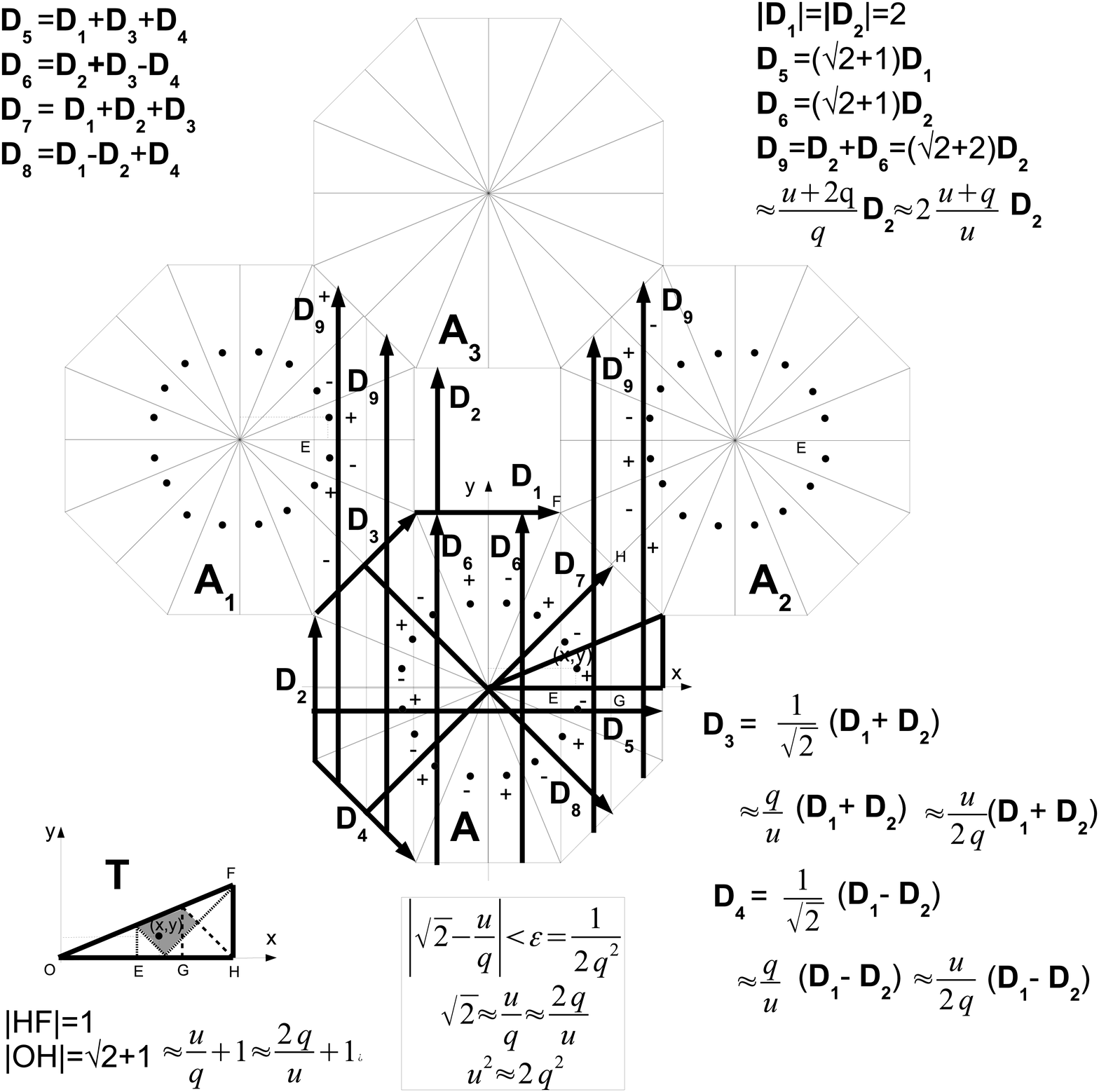,width=7 cm} \caption{The Bogomolny-Schmit triangle {\bf T}, its EPP {\bf A} and the three periodic copies
${\bf A}_i,\;i=1,2,3$ of {\bf A}. The four independent periods ${\bf D}_i,\;i=1,...,4$, are shown as well as the two POCs with the period ${\bf D}_6$
and the four POCs with the period ${\bf D}_9$. The lines EF and GH in the triangle {\bf T} are the nodal ones of the SWF \mref{16} corresponding to the
singular diagonals EF and GH on EPP. The distribution of the images of the point $(x,y)$ shown in the figure are ensured by the shaded area in the
triangle {\bf T}}
\end{center}
\end{figure}

\subsubsection{The quantization on aperiodic skeletons}

\hskip+2em Assuming the approximation as in Fig.2 (i.e. $\sqrt{2}\approx u/q$ or $\sqrt{2}=2/\sqrt{2}\approx 2q/u$ with the same $\epsilon$-accuracy) we
make the step RPB$\to$DRPB and then we can quantize the case according to section 2. Quantizing on aperiodic skeletons we get therefore two possibilities
(call them the $u$- or the $q$-approximations respectively)
\be
{\bf p}\cdot{\bf D}_1=2\pi um\;\;\;\;\;\;\;\;\;\;\;\;\nn\\
{\bf p}\cdot{\bf D}_2=2\pi un\;\;\;\;\;\;\;\;\;\;\;\;\;\nn\\
m,n=\pm 1,\pm 2,...
\label{13a}
\ee
and
\be
{\bf p}\cdot{\bf D}_1=2\pi qm\;\;\;\;\;\;\;\;\;\;\;\;\nn\\
{\bf p}\cdot{\bf D}_2=2\pi qn\;\;\;\;\;\;\;\;\;\;\;\;\;\nn\\
m,n=\pm 1,\pm 2,...
\label{13b}
\ee
where we have taken into account that according to App.A and the approximations we have done both ${\bf D}_{1,2}/u$ are the approximate periods of the
RPRS as well as ${\bf D}_{1,2}/q$. However both these approximations we have to consider separately, i.e. as the two different ones.

Note also that in both the above cases $m$ and $n$ have to be unequal zero since in any of such a case the skeleton would be periodic.

Therefore for the $u$-approximation we have
\be
{\bf p}_{mn}=2\pi u\frac{(m{\bf D}_2-n{\bf D}_1)\times({\bf D}_1\times{\bf D}_2)}{({\bf D}_1\times{\bf D}_2)^2}=
\fr\pi u(m{\bf D}_1+n{\bf D}_2)\nn\\
m,n=\pm 1,\pm 2,...
\label{14}
\ee
and a similar expression for the $q$-approximation.

For the semiclassical energy spectrum we then get
\be
E_{mn}^{(u)}=\fr{\bf p}_{mn}^2=\fr\pi^2u^2(m^2+n^2)\nn\\
m,n=\pm 1,\pm 2,...
\label{15}
\ee
and
\be
E_{mn}^{(q)}=\fr{\bf p}_{mn}^2=\fr\pi^2q^2(m^2+n^2)\nn\\
m,n=\pm 1,\pm 2,...
\label{15a}
\ee
respectively.

The spectra  \mref{15} and \mref{15a} are of course different, i.e. they approximate different parts of the energy spectra in the triangular billiards.

To get a SWF satisfying the Dirichlet boundary conditions on the triangle considered we should use \mref{13} with the $\eta$-signs shown if Fig.1 with the
following result (up to a normalization constant) for the case of the $u$-approximation
\be
\Psi_{mn}^{(u)}(x,y)=\sin(\pi umx)\sin(\pi uny)-\sin\ll(\frac{1}{\sqrt{2}}\pi um(x+y)\r)\sin\ll(\frac{1}{\sqrt{2}}\pi un(x-y)\r)+\nn\\
\sin\ll(\frac{1}{\sqrt{2}}\pi um(x-y)\r)\sin\ll(\frac{1}{\sqrt{2}}\pi un(x+y)\r)-\sin(\pi unx)\sin(\pi umy)
\label{16}
\ee

It is seen from the above formulae that to get different states $m,n$ should be limited to the positive values only and to the pairs for which
$1\leq n\leq m$.

The above SWF vanishes by its construction on the sides OF and OH of the triangle $T$ of Fig.1 and takes the following form on its side FH
\be
\Psi_{mn}^{(u)}(\sqrt{2}+1,y)=\sin(\pi um(\sqrt{2}+1))\sin(\pi uny)-\nn\\
\sin\ll(\frac{1}{\sqrt{2}}\pi um(\sqrt{2}+1+y)\r)\sin\ll(\frac{1}{\sqrt{2}}\pi un(\sqrt{2}+1-y)\r)+\nn\\
\sin\ll(\frac{1}{\sqrt{2}}\pi um(\sqrt{2}+1-y)\r)\sin\ll(\frac{1}{\sqrt{2}}\pi un(\sqrt{2}+1+y)\r)-\sin(\pi un(\sqrt{2}+1))\sin(\pi umy)
\label{17}
\ee
and of course does not vanish on the side. It vanishes however if $\sqrt{2}$ is substituted everywhere in \mref{17} by $u/q$ or by $2q/u$ respectively
which means that $\Psi_{mn}(\sqrt{2}+1,y)$ vanishes on OH with the $\epsilon$-accuracy. More precisely we have for $y\leq 1$
\be
|\Psi_{mn}^{(u)}(\sqrt{2}+1,y)|<2\pi\frac{\epsilon u}{1-\frac{\epsilon}{\sqrt 2}}(m+n)
\label{18}
\ee
so that for $\epsilon=1/u^2=$ and $m+n<<u$ we get
\be
|\Psi_{mn}^{(u)}(1,y)|<<1\nn\\
0\leq y\leq 1
\label{19}
\ee
i.e. $\Psi_{mn}^{(u)}(x,y)$ and the spectrum \mref{15} can be considered as the good semiclassical approximations of the real quantities with the approximations being
the better the closer $\sqrt{2}$ is the rational $u/q$.

Exactly in the same way can be analyzed the $q$-approximation which provides us with the formulae \mref{16}-\mref{19} where the parameter $u$ is
substituted by the $q$ one.

On the triangle {\bf T} in Fig.1 are shown also the lines $x=1,\;y=\pm(x-\sqrt{2}),\;x=1+\sqrt{2}/2$ and $y=-x+\sqrt{2}+1$ which are the nodal lines for
$\Psi_{mn}^{(u,q)}(x,y)$ with the
accuracies given by \mref{18}. These nodal lines coincide with the singular diagonals $EF$ and $GH$ of the respective POCs on Fig.1 folded into the
triangle.

\subsubsection{The quantization on periodic skeletons}

\hskip+2em Let us now use the periodic skeletons (POCs) shown in Fig.1 to make a respective quantization on them. There are six of them which cover totally
the EPP {\bf A} on the figure with the corresponding periods ${\bf D}_6(=(\sqrt{2}+1){\bf D}_2)$ or ${\bf D}_9(=(\sqrt{2}+2){\bf D}_2)$. Trying to
quantize on the POCs with the periods ${\bf D}_6$ and ${\bf D}_9$ we can check that the condition \mref{11b} for them is satisfied for all the periods
${\bf D}_i,\;i=1,...,4$ when $\sqrt{2}$ is approximated by $u/q$ or by $2q/u$ with no any restriction on the form of the triangle.

Trying to construct a SWF on the whole EPP we have to match the respective SWFs built on each POC on the boundaries of the latter using the forms \mref{2} and
\mref{6} of the SWFs. It is easy to note that this matching demands the momentum ${\bf p}$ to be the same for each POC and since $E_0$ being determined by the
period ${\bf D}_1$ is also the same for all the POCs the coefficient $A$ and $B$ in \mref{6} have to be also the same for all the POCs. Therefore,
quantizing according to the rule \mref{11} and \mref{11a} in the $u$-approximation, i.e. using the periods ${\bf D}_{1,2}/u$, we have
\be
{\bf p}\cdot{\bf D}_2=pD_2=2p=2\pi um\nn\\
{\bf p}\cdot{\bf D}_6=pD_6=(\sqrt{2}+1)pD_2\approx 2\pi (2u+q)m\nn\\
{\bf p}\cdot{\bf D}_9=pD_9=(\sqrt{2}+2)pD_2\approx 4\pi (u+q)m\nn\\
m=1,2,...
\label{20}
\ee
and
\be
E_{0n}=\fr\pi^2u^2n^2,\;\;\;\;\;\;\;n=0,\pm 1,\pm 2,...
\label{21}
\ee
so that
\be
E_{mn}^{(u)}=\fr p^2+E_{0n}=\fr\pi^2u^2(m^2+n^2)\nn\\
m=1,2,...,\;n=0,\pm 1,\pm 2,...
\label{22}
\ee
while in the $q$-approximation we get respectively
\be
E_{mn}^{(q)}=\fr p^2+E_{0n}=\fr\pi^2q^2(m^2+n^2)\nn\\
m=1,2,...,\;n=0,\pm 1,\pm 2,...
\label{21a}
\ee
i.e. we get exactly the same form of the energy spectrum for SWFs built on the periodic skeletons as in the
cases of the aperiodic ones, i.e. the results \mref{15} and \mref{15a}. Formally however the results \mref{22} and \mref{21a} are valid for $n<<m$.

For a SWF satisfying the Dirichlet boundary condition in the triangle considered we have to use again the formula \mref{13} with the form \mref{6} of
the BSWF. Doing this one can convince oneself that the cosine part of \mref{6} have to vanish when summing while the sine one gives exactly the result
\mref{16} and its $q$-variant for the aperiodic case.

It is possible of course to quantize the billiards on other periodic skeletons with the periods ${\bf D}_5$ or ${\bf D}_7$ or ${\bf D}_8$ and on
the respective parallel ones obviously
with the same results what is easily seen from the formula \mref{16} which is invariant under the rotations by the angles $\pm\pi/2$ and $\pm\pi/4$ so
that our initial choice of the periodic skeletons was not in fact some specific.

\subsubsection{The superscar states of Bogomolny and Schmit}

\hskip+2em According to Bogomolny and Schmit in each of the six POCs shown in Fig.1 should exist superscar states which could be excited if the ball
energy in the triangular billiards is close to an energy of some states of the POC considered. Comparing however the results of folding these POCs into
the billiards it is seen that such foldings are the same for the POCs with the same periods. Therefore it is enough to consider only two POCs with the
two different periods ${\bf D}_6$ and ${\bf D}_9$.

Consider therefore the right POC defined by the period ${\bf D}_6$ with its singular diagonal coinciding with the $y$-axis and the other one coinciding
with the line $EF$. According to
the author mentioned the respective energy spectrum and the corresponding superscar wave function satisfying the Dirichlet condition on the POC
boundaries are
\be
E_{6,mn}^{(B-S)}=\fr\pi^2\ll(m^2+\frac{n^2}{(\sqrt{2}+1)^2}\r)=\fr\pi^2(m^2+n^2(\sqrt{2}-1)^2)\nn\\
|m|,|n|>1
\label{22a}
\ee
and
\be
\Psi_{6,mn}^{(B-S)}(x,y)=A\sin\ll(\frac{1}{\sqrt{2}+1}\pi ny\r)\sin(\pi mx)+B\cos\ll(\frac{1}{\sqrt{2}+1}\pi ny\r)\sin(\pi mx)\nn\\
0\leq x\leq 1,\;\;\;\;|m|,|n|\geq 1
\label{22b}
\ee

The above wave function should be next inscribed into the triangle by the folding operation which means in fact a coherent sum of the above
B-S wave function taken at the points of the EPS of Fig.1 lying in the POC considered providing (up to a normalization constant) the following wave
function
\be
\Psi_{6,mn}(x,y)=\nn\\
-\sin((\sqrt{2}-1)\pi nx)\sin(\pi my)+\sin\ll(\pi n\frac{1}{2+\sqrt{2}}(x+y)\r)\sin\ll(\pi m\frac{\sqrt{2}}{2}(x-y)\r)\nn\\
|m|,|n|\geq 1
\label{22c}
\ee
where the point $(x,y)$ is in the shaded area of the triangle {\bf T} in Fig.1.

The above wave function satisfy the Dirichlet boundary conditions on the triangular billiards boundary.

Similarly, for the POC next to the one just considered and defined by the period ${\bf D}_9$ we have for the energy spectrum
\be
E_{9,mn}=\frac{\pi^2}{2}\ll(2m^2+\frac{n^2}{(\sqrt{2}+2)^2}\r)\nn\\
|m|,|n|>1
\label{22d}
\ee
while the corresponding SWF has the following form
\be
\Psi_{9,mn}(x,y)=\nn\\
\sin(\pi \sqrt{2}m(x-1))\sin\ll(\fr\pi n(2-\sqrt{2})y\r)-
\sin\ll(\fr\pi n(\sqrt{2}-1)(x-y)\r)\sin(\pi m(x+y-\sqrt{2}))\nn\\
|m|,|n|\geq 1
\label{22e}
\ee
where again the point $(x,y)$ is in the shaded area of the triangle {\bf T} in Fig.1.

It is important however to notice at this moment that both the superscar states $\Psi_{6,mn}(x,y)$ and $\Psi_{9,mn}(x,y)$
are discontinuous inside the
triangular billiards on the lines which are images of the singular diagonals of the POCs considered (shown in the triangle {\bf T} in Fig.1) breaking in
this way the basic demand for every properly constructed quantum-mechanical wave function.

Further, forms of the POC wave functions are not fixed and depend on distributions between POCs of all mirror images of a billiards point obtained by
unfolding the polygon billiards into its EPP. It is just this dependence which generates the discontinuity of the POC wave functions on the singular
diagonals when a mirror image of a billiards point enters or leaves a given POC crossing a singular diagonal being one of its two boundaries.

To compare the solutions $\Psi_{6,mn}(x,y)$ and $\Psi_{9,mn}(x,y)$ with the solutions $\Psi_{mn}^{(u)}(x,y)$ and $\Psi_{mn}^{(q)}(x,y)$ let us rewrite
them in their full semiclassical approximations, i.e. by substituting everywhere $\sqrt{2}$ by $u/q$ or $2q/u$. We get respectively
\be
\Psi_{mn}^{(u)}(x,y)=\sin(\pi umx)\sin(\pi uny)-\sin(\pi qm(x+y))\sin(\pi qn(x-y))+\nn\\
\sin(\pi qm(x-y))\sin(\pi qn(x+y))-\sin(\pi unx)\sin(\pi umy)\nn\\
\Psi_{mn}^{(q)}(x,y)=\sin(\pi qmx)\sin(\pi qny)-\sin\ll(\fr\pi um(x+y)\r)\sin\ll(\fr\pi un(x-y)\r)+\nn\\
\sin\ll(\fr\pi um(x-y)\r)\sin\ll(\fr\pi un(x+y)\r)-\sin(\pi qnx)\sin(\pi qmy)\nn\\
\Psi_{6,m'n'}(x,y)=\nn\\
\label{23b}
\sin\ll(\pi\frac{q}{u}m'(x-y)\r)\sin\ll(\pi\frac{q}{2q+u}n'(x+y)\r)-
\sin\ll(\pi\frac{u-q}{q}n'x\r)\sin(\pi m'y)=\\
\label{23c}
\sin\ll(\pi\frac{u}{2q}m'(x-y)\r)\sin\ll(\pi\frac{u}{2(q+u)}n'(x+y)\r)-
\sin\ll(\pi\frac{2q-u}{u}n'x\r)\sin(\pi m'y)\\
\Psi_{9,m'n'}(x,y)=\nn\\
\label{23d}
\sin\ll(\pi\frac{2q}{u}m'(x-1)\r)\sin\ll(\pi\frac{q}{2q+u}n'y\r)-\nn\\
\sin\ll(\pi\frac{u-q}{2q}n'(x-y)\r)\sin(\pi m'(x+y-2q/u))=\\
\label{22f}
\sin\ll(\pi\frac{u}{q}m'(x-1)\r)\sin\ll(\pi\frac{u}{2(q+u)}n'y\r)-\nn\\
\sin\ll(\pi\frac{2q-u}{2u}n'(x-y)\r)\sin(\pi m'(x+y-u/q))
\ee

Now we can see that the substitution $m'=um,\;n'=(2q+u)n$ in $\Psi_{6,m'n'}(x,y)$ in \mref{23b} restores the last two terms in $\Psi_{mn}^{(u)}(x,y)$
while the substitution $m'=un,\;n'=(2q+u)m$ restores its first two ones. On the other hand the substitution $m'=qm,\;n'=(q+u)n$ in \mref{23c}
restores the last two terms in $\Psi_{mn}^{(q)}(x,y)$ while substituting $m'=qn,\;n'=(q+u)m$ restores its first two ones.

Similarly, the substitution $2m'=um,\;n'=(2q+u)n$ in \mref{23d} restores the first two terms of $\Psi_{mn}^{(q)}(x,y)$
while by exchanging $m$ with $n$ in the last substitution one recovers the last two terms in $\Psi_{mn}^{(q)}(x,y)$. Analogously substituting in \mref{22f}
$m'=qm,\;n'=2(q+u)n$ one recovers the first two terms of $\Psi_{mn}^{(u)}(x,y)$ and exchanging $m$ with $n$ in the last substitution one recovers the
two last ones.

Therefore the following conclusions can be drawn from our calculations we have done in this section for the triangular billiards of Bogomolny and Schmit
\begin{itemize}
\item since there are no restrictions on the triangular billiards parameters preventing the quantization on the periodic skeletons, i.e. the condition
\mref{11b} can be satisfied by properly fixing the integers $r,s$, the results of such a quantization, i.e. the energy spectrum and SWFs, are identical
with the ones performed on aperiodic skeletons;
\item the quantization on periodic skeletons shows that the SWFs are built of contributions provided by all POCs from which the respective periodic
skeletons are formed;
\item the basic difference between SWFs built on POCs in sec.4.1.2 and the Bogomolny-Schmit states of sec.4.1.3 is that the former are built of the POC
states to be smooth on the singular diagonals while the Bogomolny-Schmit states are defined in the neighbor POCs completely independently of each other,
i.e. there is a unique SWF
defined on the whole area of the triangular EPP in the former case and six independent Bogomolny-Schmit states in the latter case;
\item the Bogomolny-Schmit superscar states appear to be immanent parts of the SWFs in triangular
billiards for some particular "resonant" values of the SWF quantum numbers. These parts cannot however exist separately outside the properly
constructed SWFs in the billiards simply because they do not satisfy elementary quantum mechanical condition such as the smoothness. This unavoidable
defect of the superscar states has been also noticed by Bogomolny and Schmit themselves \cite{46};
\item the singular diagonals of the POCs shown in Fig.1 which are visible in the triangle {\bf T} as the approximate nodal lines of the SWFs \mref{16}
are not as such for the Bogomolny - Schmit states folded into the triangle, i.e. these lines can appear only as a result of the fact that the SWFs \mref{16}
is the coherent sum of contributions from all the component POCs of the periodic skeleton;
\item the superscar effect which exists in the triangular billiards in the form of the (approximate) nodal lines of the SWFs \mref{16} being a
print of the POC singular diagonals needs not to invoke such unusual states as the superscar ones which are incompatible with the elementary quantum
mechanical demands.
\end{itemize}

To support the last claim let us compare the relations between the three energy levels which exact numerical values have been cited by Bogomolny and
Schmit \cite{46} with the respective relations provided by the formulae \mref{15}-\mref{15a}. The respective exact levels are (in arbitrary units)
$E_1=407,4,\;E_2=1015,97,\;E_3=1968,97$ which give the following relations between them
\be
\frac{E_2}{E_1}=2,4937,\;\;\;\;\;\;\frac{E_3}{E_2}=1,9380
\label{22g}
\ee

The above levels have been compared by Bogomolny and Schmit with the levels $E_{6,mn}^{(B-S)}$ for $m=50,n=1$, $m=79,n=1$ and $m=110,n=1$ respectively
which correspond to the levels $E_{mn}^{(u)}$ of the formula \mref{15} with $m=121,n=1$, $m=191,n=1$ and $m=266,n=1$. To get this correspondence we have
approximated $\sqrt{2}$ by its continued fraction to get $3363/2378$ as its approximation (so that $|\sqrt{2}-3363/2378|<1/3363^2$) since then the
sums $m+n$ for the levels considered are much less than each of the numbers $3363,\;2378$ allowing to satisfy the condition \mref{19}. Then we have
\be
\frac{E_{191,1}^{(u)}}{E_{121,1}^{(u)}}=2,4916,\;\;\;\;\;\;\frac{E_{266,1}^{(u)}}{E_{191,1}^{(u)}}=1,9395
\label{22h}
\ee

The results \mref{22h} compared with the relations \mref{22g} show not only that the superscar idea of Bogomolny and Schmit is unnecessary to explain
the superscar phenomenon but demonstrate also the accuracy of the semiclassical approximation method formulated in sec.2.

In the next few subsections we consider further examples of RPBs supporting our last conclusions.

\subsection{The parallelogram billiards}

\hskip+2em  Consider now the parallelogram billiards shown in Fig.2. As it follows from the figure an arrangement of the singular diagonals defining
boundaries of the respective POCs is repeatable with each increase of the length of the side $L$ of the parallelogram by three units. Therefore not loosing a generality
of our considerations we can limit them to some of the parallelograms shown in Fig.2. For a simplicity we will choose the parallelogram with $L=4$. It
defines of course a DRPB.
\begin{figure}
\begin{center}
\psfig{figure=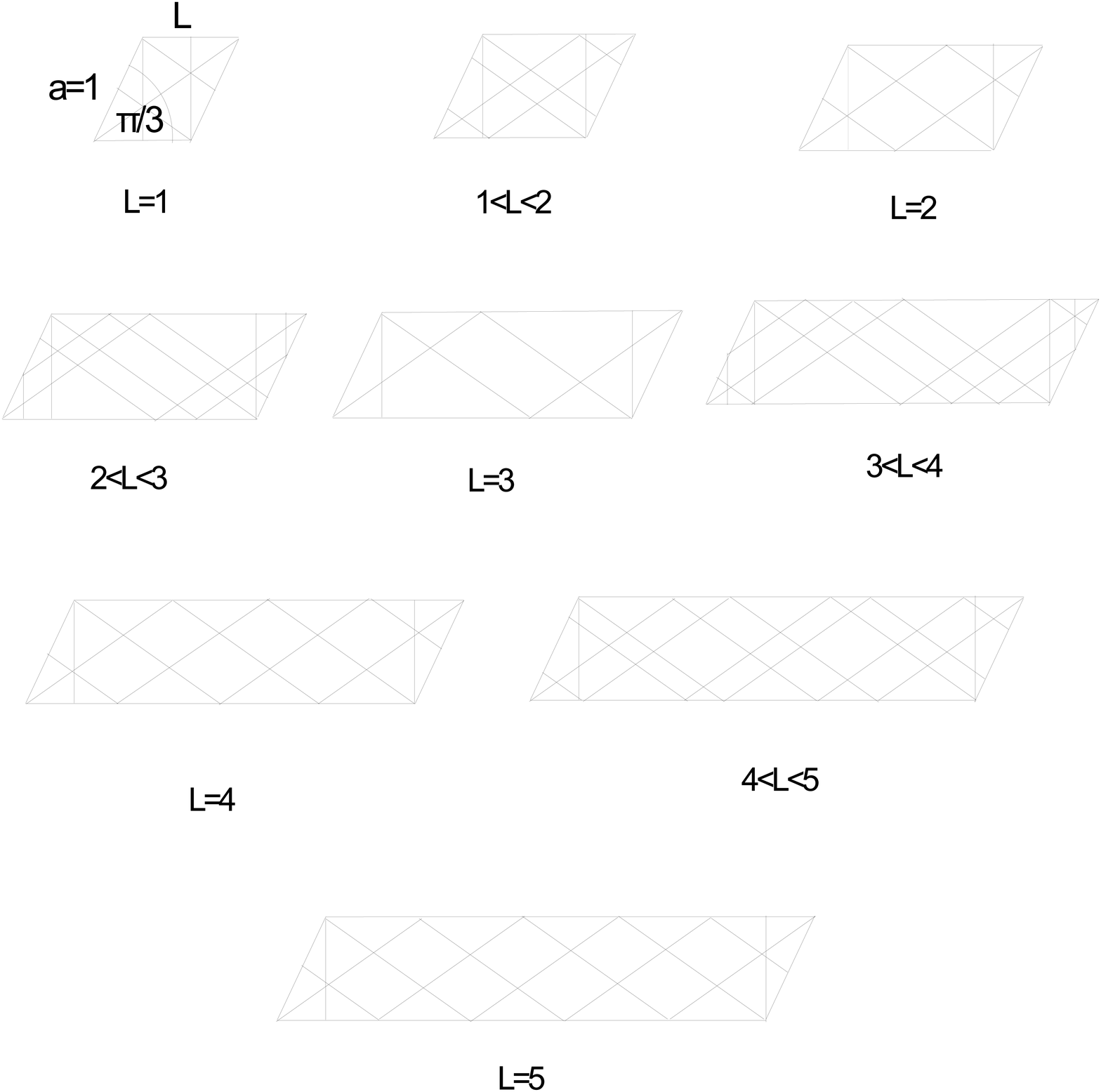,width=9 cm} \caption{The runs of the singular diagonals of POCs with their periods parallel to ${\bf d}_8$ (see Fig.3)
in the parallelograms differing by a length of their side $L$. A repetition of the pattern of SD for each increase of $L$ by three units can be
observed}
\end{center}
\end{figure}

\begin{figure}
\begin{center}
\psfig{figure=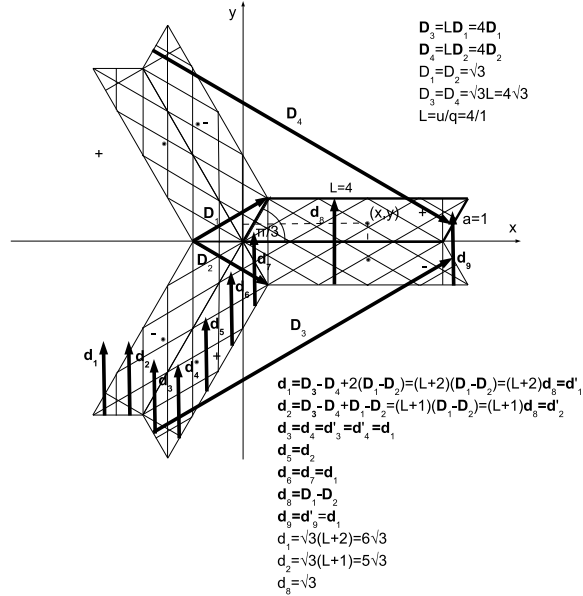,width=8 cm} \caption{The EPP of the parallelogram billiards for $L=4$ and its four independent periods ${\bf D}_i,\;i=1,...,4$, together
with the periods ${\bf d}_i,\;i=1,...,9$ and ${\bf d}'_i,\;i=1,2,3,4,9$, (which are not shown) defining the respective POCs. The periods which are not
shown are the ones of the POCs being a mirror reflection in the $x$-axis of the POCs with the periods ${\bf d}_i,\;i=1,2,3,4,9$}
\end{center}
\end{figure}

\subsubsection{The quantization on aperiodic skeletons}

\hskip+2em Choosing the periods ${\bf D}_i,\;i=1,...,4$, as the independent ones it is seen that conditions quantizing possible momenta are the
following
\be
{\bf p}\cdot{\bf D}_1=\frac{3}{2}p_x+\frac{\sqrt{3}}{2}p_y=2\pi qm\;\;\;\;\;\;\;\;\;\;\;\;\nn\\
{\bf p}\cdot{\bf D}_2=\frac{3}{2}p_x-\frac{\sqrt{3}}{2}p_y=2\pi qn\;\;\;\;\;\;\;\;\;\;\;\;\;\nn\\
m,n=0,\pm 1,\pm 2,...,\;\;\;\;|m|+|n|>0
\label{25}
\ee
so that
\be
{\bf p}_{x,mn}=\frac{2\pi}{3}(m+n)q\nn\\
{\bf p}_{y,mn}=\frac{2\pi}{\sqrt{3}}(m-n)q\nn\\
m,n=0,\pm 1,\pm 2,...,\;\;\;\;|m|+|n|>0
\label{26}
\ee
and for the semiclassical energy spectrum we get
\be
E_{mn}=\fr{\bf p}_{mn}^2=\pi^2q^2\ll(\frac{2}{9}(m+n)^2+\frac{2}{3}(m-n)^2\r)=\frac{8\pi^2q^2}{9}(m^2-mn+n^2)\nn\\
m,n=0,\pm 1,\pm 2,...,\;\;\;\;|m|+|n|>0
\label{27}
\ee
while there are two corresponding SWFs for each level, i.e. each level is at least doubly degenerate. Namely, we have first the following complex solution
\be
\Psi_{mn}(x,y)=\nn\\
e^{-\frac{\pi i}{3}(m+n)q(x+\sqrt{3}y)}\sin\ll(\frac{\pi q}{\sqrt{3}}(m-n)(\sqrt{3}x-y)\r)-\nn\\
e^{-\frac{\pi i}{3}(m+n)q(x-\sqrt{3}y)}\sin\ll(\frac{\pi q}{\sqrt{3}}(m-n)(\sqrt{3}x+y)\r)+
e^{\frac{2\pi i}{3}(m+n)qx}\sin\ll(\frac{2\pi}{\sqrt{3}}(m-n)qy\r)
\label{28}
\ee
which real and imaginary parts provide us with the two real SWFs, namely
\be
\Psi_{mn}^{(1)}(x,y)=\nn\\
-\sin\ll(\frac{\pi}{3}(m+n)q(x+\sqrt{3}y)\r)\sin\ll(\frac{\pi q}{\sqrt{3}}(m-n)(\sqrt{3}x-y)\r)+\nn\\
\sin\ll(\frac{\pi}{3}(m+n)q(x-\sqrt{3}y)\r)\sin\ll(\frac{\pi q}{\sqrt{3}}(m-n)(\sqrt{3}x+y)\r)+\nn\\
\sin\ll(\frac{2\pi}{3}(m+n)qx\r)\sin\ll(\frac{2\pi}{\sqrt{3}}(m-n)qy\r)\nn\\
\Psi_{mn}^{(2)}(x,y)=\nn\\
\cos\ll(\frac{\pi}{3}(m+n)q(x+\sqrt{3}y)\r)\sin\ll(\frac{\pi q}{\sqrt{3}}(m-n)(\sqrt{3}x-y)\r)-\nn\\
\cos\ll(\frac{\pi}{3}(m+n)q(x-\sqrt{3}y)\r)\sin\ll(\frac{\pi q}{\sqrt{3}}(m-n)(\sqrt{3}x+y)\r)+\nn\\
\cos\ll(\frac{2\pi}{3}(m+n)qx\r)\sin\ll(\frac{2\pi}{\sqrt{3}}(m-n)qy\r)\nn\\
\label{29}
\ee

It is worth to note that the solutions \mref{29} are odd with respect to the variable $y$ to satisfy the Dirichlet boundary conditions we have
demanded. The respective even solutions would correspond to the Neumann boundary conditions.

However one can check also that both the above solutions vanish on the line $y=-\sqrt{3}(x-1)$ which the property prevents in fact the existence of SWFs
which are even with respect to the mirror reflection by this line in the case when $L=1$, i.e. when the parallelogram reduces to the rhombus what was
first observed by Richens and Berry \cite{53}. This is an illustration of the limitations of our approach to the semiclassical description of the quantum
states in RPB showing that not all of these states can be catch by the method.

\subsubsection{The quantization on periodic skeletons}

\hskip+2em Consider now the quantization on periodic skeletons for which we have chosen those seen in Fig.3, i.e. parallel to the period ${\bf d}_8/q$.
There are fourteen of them in the figure having periods shown in the figure in the respective details. The condition \mref{11b} can be satisfied by each
POC of the periodic skeleton chosen by taking the period ${\bf D}_1/q$ as the second one in the respective quantization formulae \mref{11}-\mref{12}
which adjusts the respective integers $r,s$ in \mref{17} on the values $r=1,\;s=2$, i.e. there are no any restriction on the parameters of the
parallelogram. Therefore the condition \mref{11b} is satisfied for any other period of EPP of Fig.3.

Taking further the respective BSWF in the form
\be
\Psi_{BSWF}(x,y)=e^{ipy}(A\cos(\sqrt{2E_0}x)+B\sin(\sqrt{2E_0}x))
\label{29a}
\ee
and enforcing its periodicity on the periods ${\bf D}_i/q,\;i=1,2$ we get
\be
\frac{3}{2}\sqrt{2E_0}+\frac{\sqrt{3}}{2}p=2\pi qm\;\;\;\;\;\;\;\;\;\;\;\;\nn\\
\frac{3}{2}\sqrt{2E_0}-\frac{\sqrt{3}}{2}p=2\pi qn\;\;\;\;\;\;\;\;\;\;\;\;\;\nn\\
m,n=0,\pm 1,\pm 2,...,\;\;\;\;m>0
\label{29b}
\ee
and hence
\be
p=\frac{2}{\sqrt{3}}\pi(m-n)q\nn\\
E_0=\frac{2}{9}\pi^2(m+n)^2q\nn\\
m\geq |n|,\;n=0,\pm 1,\pm 2,...,
\label{30}
\ee
so that the corresponding energy spectrum is
\be
E_{mn}=\pi^2q^2\ll(\frac{2}{3}(m-n)^2+\frac{2}{9}(m+n)^2\r)
\label{31}
\ee

Comparing the last formula with the previous one \mref{27} it is seen that their forms coincide.

Constructing now SWFs corresponding to the periodic skeletons considered we have to start from the BSWF \mref{29a}
to make a coherent sum of them over the points shown in Fig.3. Then the term with the coefficient
$A$ in \mref{29a} will reproduce $\Psi_{mn}^{(1)}(x,y)$ in \mref{29} while this with the coefficient $B$ - $\Psi_{mn}^{(2)}(x,y)$ in \mref{29}.

\subsubsection{The superscar states of Bogomolny and Schmit}

\hskip+2em The Bogomolny-Schmit superscar states can be built in every POC. However a number of POCs in an EPP constructed for a given RPB can in general depend
on lengths of its sizes which can be changed while all angles of RPB are frozen \cite{41}. The considered case is just of this kind while
the Bogomolny-Schmit triangle considered previously is completely insensitive on such changes. Therefore a number of POCs shown in Fig.3 corresponds
just to the length $L$ actually shown in the figure which is equal to 4. There are three type of POCs parallel to the $y$-axis differing
by their periods and wides. One type of them has the period  ${\bf d}_3$ and the wide $1/2$, another the period ${\bf d}_5$ and also the wide $1/2$ while
the third one - the bouncing ball type - has the period ${\bf d}_8$ and the wide $7/2$.

Constructing the Bogomolny-Schmit superscar states let us note that the states built in the POCs with the same periods are the same when inscribed into
the parallelogram. Therefore it is enough to build them for the POCs with the periods ${\bf d}_3,\;{\bf d}_5$ and ${\bf d}_8$ only. We get then for the
respective energy spectra
\be
E_{3,m'n'}^{(B-S)}=E_{3',m'n'}^{(B-S)}=2\pi^2\ll(\frac{m'^2}{3(L+2)^2}+n'^2\r)=2\pi^2\ll(\frac{m'^2q^2}{3(u+2q)^2}+n'^2\r)\nn\\
E_{5,m'n'}^{(B-S)}=2\pi^2\ll(\frac{m'^2}{3(L+1)^2}+n'^2\r)=2\pi^2\ll(\frac{m'^2q^2}{3(u+q)^2}+n'^2\r)\nn\\
E_{8,m'n'}^{(B-S)}=2\pi^2\ll(\frac{1}{3}m'^2+\frac{n'^2}{\ll(2L-1\r)^2}\r)=2\pi^2\ll(\frac{1}{3}m'^2+\frac{n'^2q^2}{(2u-q)^2}\r)\nn\\
|m'|,|n'|\geq 1
\label{34}
\ee
while the corresponding superscar wave functions are (up to normalization factors)
\be
\Psi_{3,m'n'}(x,y)=\sin(\pi n'(x+\sqrt{3}y-L))\sin\ll(\frac{\pi m'}{\sqrt{3}(L+2)}(\sqrt{3}x-y+\sqrt{3})\r)=\nn\\
\sin\ll(\pi n'(x+\sqrt{3}y-\frac{u}{q})\r)\sin\ll(\frac{\pi qm'}{\sqrt{3}(u+2q)}(\sqrt{3}x-y+\sqrt{3})\r)\nn\\
\Psi_{5,m'n'}(x,y)=\sin(\pi n'(x-\sqrt{3}y))\sin\ll(\frac{\pi m'}{\sqrt{3}(L+1)}(\sqrt{3}x+y)\r)=\nn\\
\sin(\pi n'(x-\sqrt{3}y))\sin\ll(\frac{\pi qm'}{\sqrt{3}(u+q)}(\sqrt{3}x+y)\r)\nn\\
\Psi_{8,m'n'}(x,y)=\sin\frac{\pi n'(x-\fr)}{L-\fr}\sin\ll(\frac{2}{\sqrt{3}}\pi m'y\r)=
\sin\frac{2\pi n'q(x-\fr)}{2u-q}\sin\ll(\frac{2}{\sqrt{3}}\pi m'y\r)\nn\\
|m'|,|n'|\geq 1
\label{35}
\ee

Comparing the above superscar solutions with the SWFs \mref{29} one can easily identify them with the respective terms of $\Psi_{mn}^{(1)}(x,y)$ in
\mref{29}. Namely, putting $m+n=3m'',\;m-n=n''$ in \mref{29} and subsequently $m'=(u+2q)n'',\;n'=m''q$ in $\Psi_{3,m'n'}(x,y)$,
$m'=(u+q)n'',\;n'=m''q$ in $\Psi_{5,m'n'}(x,y)$ and $m'=n''q,\;n'=m''(2u-q)$ in $\Psi_{8,m'n'}(x,y)$ identifies the latter
supperscar states and their spectra with the first, second and third terms of $\Psi_{mn}^{(1)}(x,y)$ respectively as well and the spectrum \mref{27} with
the respective spectra \mref{34}.

Therefore one can repeat here the final conclusions of the previous section for the triangular billiards.

\subsection{POCs in the rectangle - the superscar phantoms}

\hskip+2em This case seems to be trivial but it has the following properties distinguishing it from the previous cases considered
\begin{itemize}
\item a quantization on a periodic skeleton depends on the sizes of the rectangle which have to be different for different POCs;
\item all POCs in the rectangular billiards can be identified despite the fact that their number is infinite; and
\item an exceptional regular periodic structure of all POCs having the same period on the corresponding RPRS generate an additional "dynamical" period of
the semiclassical wave functions which cannot be obtained from the ones determining the corresponding RPRS;
\item it is a rare example of the RPB which all the semiclassical eigen states are the superscar states of Bogomolny and Schmit.
\end{itemize}

In our earlier paper \cite{41} stimulated in fact by the Bogomolny-Schmit one \cite{46} we have shown that POCs are common in PBs giving rise to study
possible periodic supperscar SWFs propagating along such POCs. In particular we have identified an infinite number of such POCs and the corresponding
superscars in the rectangle. However their existence in the rectangle seemed to be unreal as solutions since they did not satisfy
the typical quantum condition of smoothness contrary to the well known simple solution to the quantized rectangle satisfying
the condition mentioned which is both exact and semiclassical and having the well known form
\be
\Psi_{mn}(x,y)=A\sin(\pi m\frac{x}{a})\sin(\pi n\frac{y}{b}), \;\;\;\;\;m,n=1,2,...
\label{23}
\ee
with the energy spectrum
\be
E_{mn}=\frac{\pi^2}{2}\ll(\frac{m^2}{a^2}+\frac{n^2}{b^2}\r)
\label{24}
\ee
where $a,b$ are the side lengths of the rectangle (see Fig.4).
\begin{figure}
\begin{center}
\psfig{figure=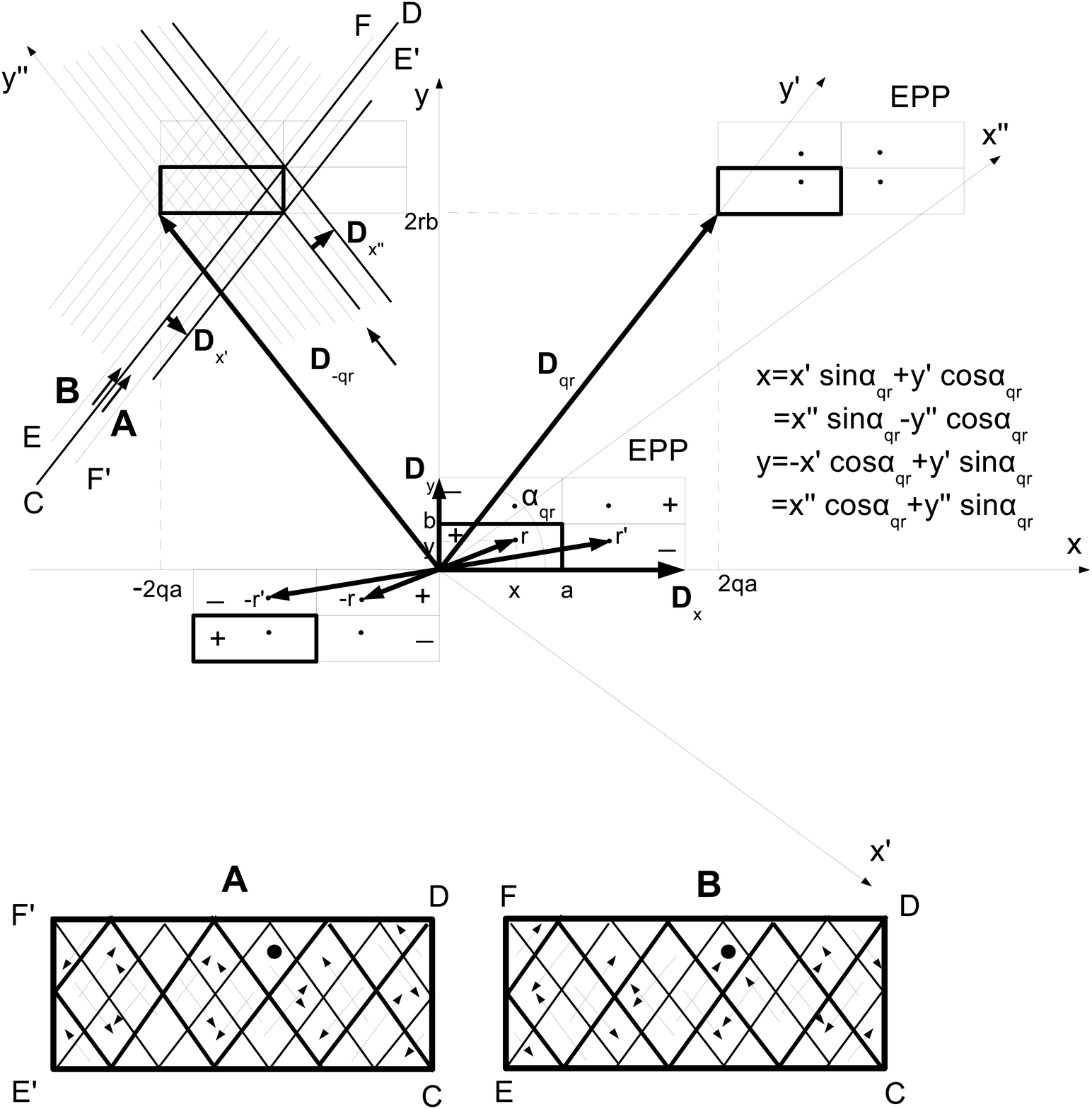,width=8cm} \caption{The EPP of the rectangular billiards and its two independent periods ${\bf D}_i,\;i=x,y$ together with the
periods ${\bf D}_{qr}$ and ${\bf D}_{-qr}$ defining the respective POCs each of the width $(a/r)\sin\alpha_{qr}$. The figures {\bf A} and {\bf B}
show the runnings of the singular diagonals $EF$, $CD$ and $F'E'$ of the POCs {\bf A} and {\bf B} folded into the rectangle.}
\end{center}
\end{figure}

In fact the solutions constructed in our paper \cite{46} according to the Bogomolny-Schmit rules were discontinuous inside the rectangle differing also on
the first glance by their energy spectra from \mref{24}. We will show below that as in the previous cases considered above such states cannot exist in
the rectangular billiards contrary to the superscar phenomena themselves which existence is ensured by the immanent structure of the semiclassical wave functions
in the billiards.

\subsubsection{Quantization on aperiodic skeletons}

\hskip+2em The periodic structure of RPRS defined by the rectangular billiards in Fig.4 is governed of course by the two independent periods
${\bf D}_x=[2a,0]$ and ${\bf D}_y=[0,2b]$ of the EPP. We can choose them to quantize the rectangular billiards semiclassically, namely
\be
{\bf p}\cdot{\bf D}_x=2p_xa=2\pi m''\nn\\
{\bf p}\cdot{\bf D}_y=2p_yb=2\pi n''\nn\\
m'',n''=0,\pm 1,\pm 2,...,\;\;\;|m''|+|n''|>0
\label{117}
\ee
and then the energy spectrum for the considered case of the rectangle can have also the form typical for the aperiodic case
\be
E_{m'',n''}=\fr(p_x^2+p_y^2)=\frac{\pi^2}{2}\ll(\frac{m''^2}{a^2}+\frac{n''^2}{b^2}\r)
\label{118}
\ee
which coincides with the exact formula \mref{24} while the corresponding SWF is determined by \mref{13} which according to Fig.4 gives
\be
\Psi_{m''n''}^{(sem)}(x,y)=A'\ll(e^{i{\bf p}{\bf r}}+e^{-i{\bf p}{\bf r}}-e^{i{\bf p}{\bf r}'}-e^{-i{\bf p}{\bf r}'}\r)=
-4A'\sin(\pi m''\frac{x}{a})\sin(\pi n''\frac{y}{b})
\label{119}
\ee
i.e. it coincides with \mref{23}.

The above coincidences are well known facts of the semiclassical treatment of the rectangular billiards.

\subsubsection{Quantization on periodic skeletons}

\hskip+2em Consider now the periodic skeletons in the rectangular billiards which EPP is shown in Fig.4 defined by two coprime positive integers $q,r$ which periodic
trajectories running through the plane RPRS corresponding to the case make the angel $\alpha_{qr}$ with the $x$-axis with $\tan\alpha_{qr}=rb/qa$ where $a,b$ are
lengths of the respective sides of the rectangle. The period ${\bf D}_{qr}$ of the skeleton is then equal to ${\bf D}_{qr}=[2qa,2rb]=q{\bf D}_x+
r{\bf D}_y$ with the length
$D_{qr}=2\sqrt{q^2a^2+r^2b^2}$. The singular diagonal corresponding to the case which crosses the point $(0,0)$ on Fig.4 crosses also the point $(2qa,2rb)$ of
the RPRS. A SD defined by the pair $r,q$ bounces $r-1$-times from each horizontal side of the rectangle and $q-1$-times - from
each of the vertical ones. Pairs $r,q$ can appear in the following combinations: $(e,o)$, $(o,o)$ and $(o,e)$ where $e$ stands for "even"
and $o$ - for "odd" numbers. The respective SDs defined by these combinations finish their runs through the rectangle in the vertices $(a,0)$, $(a,b)$
and $(0,b)$ correspondingly if each of them starts at the point $(0,0)$ of the RPRS (see Fig.4). Of course their total lengths are equal only the half
of the period $D_{qr}$.

Quantizing the rectangular billiards on the parallel periodic skeletons (POCs) having each the period ${\bf D}_{qr}$ we have for the quantum ball in each
POC
\be
{\bf p}_m\cdot{\bf D}_{qr}=p_mD_{qr}=2\pi m,\;\;\;\;\;\;\;m=0,1,2,...
\label{111}
\ee
where we have assumed that the momentum ${\bf p}_m$ has the direction of ${\bf D}_{qr}$.

Assuming further the arrangement of coordinates as in Fig.4 we have for the respective BSWFs
\be
\Psi_{\pm m}^{BSWF}(x',y')=e^{\pm ip_my'}\ll(A\sin\ll(\sqrt{2E_{qr,0}}x'\r)+B\cos\ll(\sqrt{2E_{qr,0}}x'\r)\r)
\label{112}
\ee
defined in each POC with the period ${\bf D}_{qr}$.

The two periodic motions with $(q,r)=(1,0)$ and $(q,r)=(0,1)$ are particularly simple realizing the so called bouncing ball cases parallel to the
$x$-axis in the first case and to the $y$-axis - in the second one. The corresponding SWFs for these cases can be got immediately from \mref{23} and from
\mref{119} as well by the following obvious decompositions of $\Psi_{mn}(x,y)$ and $\Psi_{mn}^{(sem)}(x,y)$
\be
\Psi_{mn}^{(sem)}(x,y)\equiv\Psi_{mn}(x,y)=2iA'\ll(e^{\pi m\frac{x}{a}}\sin(\pi n\frac{y}{b})-e^{-\pi m\frac{x}{a}}\sin(\pi n\frac{y}{b})\r)=\nn\\
2iA'\ll(e^{\pi n\frac{y}{b}}\sin(\pi m\frac{x}{a})-e^{-\pi n\frac{y}{b}}\sin(\pi m\frac{x}{a})\r), \;\;\;\;\;m,n=1,2,...
\label{23a}
\ee
in which we recognize the structures obtained from \mref{112} by the rules \mref{13}. The first decomposition in \mref{23} corresponds to the quantum bouncing ball
motion parallel to the $x$-axis while the second one - along the $y$-axis. Clearly the energy spectra for both the motions are also the same and equal to
\mref{24}.

The BSWFs \mref{112} have to be periodic also with respect to the periods ${\bf D}_x$ and ${\bf D}_y$ which leads to the following conditions (note that
$p_my'={\bf p}_m\cdot{\bf r}$)
\be
{\bf p}_m\cdot{\bf D}_x=2p_{m,x}a=2\pi m'\nn\\
{\bf p}_m\cdot{\bf D}_y=2p_{m,y}b=2\pi n'\nn\\
\sqrt{2E_{qr,0}}2a\sin\alpha_{qr}=2\pi n\nn\\
m',n',n=0,1,2,...
\label{113}
\ee
so that $m=qm'+rn'$ while the side lengths $a,b$ of the rectangle have to satisfy the following condition
\be
\frac{b^2}{a^2}=\frac{n'q}{m'r}=\frac{lq}{kr}
\label{114}
\ee
where $(q,r)\neq(0,1),(1,0)$ and $k,l\neq 0$ are coprime integers so that $m'=ck,\;n'=cl$ with an integer $c\neq 0$ and
\be
m=c(kq+lr)\nn\\
c=1,2,...
\label{114a}
\ee

Note that for $(q,r)$ equal to $(0,1)$ or $(1,0)$, i.e. for the bouncing ball cases, none a constraint is put on the sizes $a,b$ of the
rectangle. In the remaining cases
the condition \mref{114} shows that the sizes $a,b$ of the rectangle cannot be arbitrary having their ratio $b/a$ restricted by the integers $k,l,q,r$
if the period ${\bf D}_{qr}$ has been fixed by the choice of $q,r$. The condition follows also of course from \mref{11b}. We assume further that the
ratio $b/a$ satisfies the condition \mref{114} for some integers $l,q,k,r$.

To get the energy spectrum for the quantum ball moving in the whole billiards area we have to match all the POC SWFs having the forms given
by \mref{112} at the boundaries of POCs. Doing this we see that the necessary conditions of the matching to be satisfied by
${\bf p}_m$ defined in each skeleton are that they must be all equal as well as all the respective $E_{qr,0}$. It means that the form \mref{112} is valid
for the whole rectangular Riemann surface. Having the BSWF matched in this way we can use it to construct the final SWF satisfying allowed boundary
conditions applying the rule \mref{13} of sec.2.2. If the boundary conditions are the Dirichlet ones then we should arrive at the SWF given by \mref{119}.

However up to this moment we have not taken into account the periodic structure of the rectangular RPRS formed by the POCs considered which certainly
has to have some influence on the periodic properties of the respective SWFs different than the one generated by the rectangular EPP. Let us
therefore discuss in some details this possible effect below.

First let us note that the BSWF \mref{112} matched in the way described above represents the running wave function in every POC with the momentum {\bf p}
shown in Fig.4. If we choose the POC denoted in Fig.4 by {\bf A} and fold it together with the BSWF defined in it into the rectangle then we get the
pattern {\bf A} shown in Fig.4 and having the following properties
\begin{enumerate}
\item  the rectangle is filled by the POC completely when folding with the continuous lines
in the figure denoting SDs corresponding to the POC while the arrows show directions of the propagation of the BSWF defined on the folded POC;
\item some folded pieces of the POC conserve their original directions parallel to the $y'$-axis or change them to the opposite one while the others
get the direction of the $y''$-axis or the opposite one;
\item each point of the rectangle is crossed twice by the folded POC;
\item every of the two SDs of the folded POC is mapped twice into the rectangle linking two different pairs of its vertexes;
\item in each pair of the two neighbour parallel pieces of the folded POC contacting on its SDs the BSWF propagates in the opposite directions;

The SWF got in the rectangle by the last two points of the above construction are in general discontinuous on the images of the SDs. Moreover it is also
still running wave, i.e. not a
standing one. To get the latter it is necessary to interfere in the rectangle two running waves with the opposite propagations. In the rectangle {\bf B} in Fig.4 there is
shown a pattern of the POC {\bf B} closest to {\bf A} folded into the rectangle with the respective BSWF which satisfies the condition of the opposite
propagation to the one defined in the POC {\bf A}. Note however that in the EPP the BSWF in both the POCs run in the same direction as it is shown on
Fig.4.

Now we can continue the construction of the SWF with the desired properties as follows
\item fold both the POCs {\bf A} and {\bf B} simultaneously into the rectangle, i.e. put the patterns {\bf A} and {\bf B} on each other so that each
point of the rectangle is covered four times by the folded POCs and therefore in each point of the rectangle there are four values of the respective
BSWF which can interfere in it;
\item the BSWF mentioned is then matched already on the common SD of the POCs {\bf A} and {\bf B}, i.e. on the line $CD$ crossing the EPP of the rectangle or equivalently
on the bold line $CD$ linking the vertexes $C$ and $D$ of the rectangles {\bf A} and {\bf B} on Fig.4;

The standing wave function obtained in this way can however be still discontinuous on the common emages of the second SDs of the POCs considered. This
discontinuity can be removed however by the following step
\item identify the BSWF on the SDs $EF$ and $F'E'$ of Fig.4 by making it periodic on the rectangular PBRS with the
period ${\bf D}_{x'}=[2a\sin\alpha_{qr}/r,0]'$, i.e. parallel to the $x'$-axis with the length equal to $(2a/r)\sin\alpha_{qr}$; then
\item we get a SWF $\Psi_{m'n'}^{(sem)}(x,y)$ satisfying the Dirichlet conditions on the rectangle sides attaching to the BSWF built in the previous
point the plus sign in the POC pieces having the directions of the $y'$-axis and the minus sign to the BSWF defined in the pieces with the $y''$-axis
directions and taking the sum of its four values in each point of the rectangle;
\end{enumerate}

It is clear therefore that the last condition in \mref{113} has to be corrected by
\be
\sqrt{2E_{qr,0;n}}\frac{2a}{r}\sin\alpha_{qr}=2\pi n\nn\\
n=0,1,2,...
\label{116}
\ee
so that the energy spectrum corresponding to the periodic skeleton just considered is
\be
E_{mn}=\fr p_m^2+E_{qr,0;n}=\fr \frac{\pi^2m^2}{q^2a^2+r^2b^2}+\frac{\pi^2n^2}{2}\frac{q^2a^2+r^2b^2}{a^2b^2}\nn\\
m=1,2,...,\;n=0,\pm 1,\pm 2,...
\label{116a}
\ee

The above discussion shows also that we have to include into our considerations also the twin periodic skeletons shown in Fig.4 with the period
${\bf D}_{-qr}$ each since their respective POCs and the BSWFs defined on them lead to the same spectrum and (up to a constant) the same SWFs in the
rectangle. In fact each such a POC when being folded into the rectangle reproduces the same pattern as the POCs with the period ${\bf D}_{qr}$ do.
Therefore we have to expect that the respective BSWF \mref{112} should have still another period ${\bf D}_{x''}$ perpendicular to ${\bf D}_{-qr}$ with
the value equal to $(2a/r)\sin\alpha_{qr}$ as it is shown in Fig.4. It will be shown below how the SWF in the rectangle billiards is built on both the
systems of the periodic skeletons.

To compare both the quantizations, i.e. on the periodic skeletons with those on the aperiodic ones let us now rewrite the spectrum \mref{118} and the
SWF \mref{119} in the coordinates $x',y'$ and  $x'',y''$ corresponding to the POCs determined by the periods
${\bf D}_{qr}$ and ${\bf D}_{-qr}$ respectively. To this goal let us project the momentum ${\bf p}=[p_x,p_y],\;p_x,p_y\geq 0$, quantized by \mref{117}
on the axes $x',y'$ to get
\be
p_{x'}={\bf p}\cdot\frac{{\bf D}_{qr}\times({\bf D}_x\times{\bf D}_{qr})}{|{\bf D}_{qr}\times({\bf D}_x\times{\bf D}_{qr})|}=
{\bf p}\cdot\frac{{\bf D}_{qr}^2{\bf D}_x-4qa^2{\bf D}_{qr}}{4rabD_{qr}}=\pi\frac{m''rb^2-n''qa^2}{ab(q^2a^2+r^2b^2)^\fr}\nn\\
p_{y'}={\bf p}\cdot\frac{{\bf D}_{qr}}{D_{qr}}=\pi\frac{qm''+rn''}{(q^2a^2+r^2b^2)^\fr}\nn\\
\label{120a}
\ee
and the momentum ${\bf p}=[-p_x,p_y],\;p_x,p_y\geq 0$, on the axes $x'',y''$ with the result
\be
p_{x''}={\bf p}\cdot\frac{{\bf D}_{-qr}\times({\bf D}_x\times{\bf D}_{-qr})}{|{\bf D}_{-qr}\times({\bf D}_x\times{\bf D}_{-qr})|}=
{\bf p}\cdot\frac{{\bf D}_{-qr}^2{\bf D}_x-4qa^2{\bf D}_{-qr}}{4rabD_{-qr}}=\nn\\
2\pi\frac{-m''rb^2+n''qa^2}{abD_{qr}}=-p_{x'}\nn\\
p_{y''}={\bf p}\cdot\frac{{\bf D}_{-qr}}{D_{-qr}}=\pi\frac{qm''+rn''}{(q^2a^2+r^2b^2)^\fr}=p_{y'}
\label{120b}
\ee

Note that in both the formulae \mref{120a} and \mref{120b} $m'',n''$ are positive both.

Therefore $E_{m'',n''}$ gets now the following form
\be
E_{m'',n''}=\fr(p_{x'}^2+p_{y'}^2)=\fr(p_{x''}^2+p_{y''}^2)=\frac{\pi^2}{2}\ll(\frac{(qm''+rn'')^2}{q^2a^2+r^2b^2}+\frac{(m''rb^2-n''qa^2)^2}{a^2b^2(q^2a^2+r^2b^2)}\r)=\nn\\
\frac{\pi^2}{2}\ll(\frac{(qm''+rn'')^2}{q^2a^2+r^2b^2}+\frac{(m''l-n''k)^2q^2a^2}{k^2b^2(q^2a^2+r^2b^2)}\r)
\label{121}
\ee

Quantizing on the aperiodic skeletons we get of course all possible quantized momenta and the full energy spectrum. Inside them must be therefore also the
momenta corresponding to the periodic skeletons which can be recovered by putting $p_{x'}=p_{x''}=0$ in \mref{120a} and \mref{120b} respectively. Then we
get $m''rb^2-n''qa^2=0$ so that according to \mref{114} $rb^2/qa^2=n''/m''=l/k$, and $m''=ck\;n''=cl,\;c=1,2,...$ . Taking into account that
$m=c(kq+lr)=m''q+n''r$ we then get from \mref{120a}-\mref{120b} for the momenta directed along the respective periods ${\bf D}_{qr}$ and ${\bf D}_{-qr}$
\be
p_{y'}^{(per)}=p_{y''}^{(per)}=2\pi\frac{m}{D_{qr}}=p_m
\label{122}
\ee

However contrary to the semiclassical quantization on the aperiodic skeletons we should also take into account that quantizing on the periodic ones
introduces the semiclassical correction $E_{qr,0;n}$ to the spectrum and the corresponding one to the SWFs as well. These corrections can be recovered
by modifying the quantum numbers $m'',n''$ putting
\be
m''=ck+nr\nn\\
n''=cl-nq\nn\\
c=1,2,...,\;n=0\pm 1,\pm 2,...
\label{123}
\ee
which does not change the momentum \mref{122} while recovers again components $p_{x'},\;p_{x''}$ to be
\be
p_{x'}^{(per)}=-p_{x''}^{(per)}=\pi\frac{nr}{a\sin\alpha_{qr}}=\frac{n}{|n|}\sqrt{2E_{qr,0;n}}
\label{123a}
\ee
i.e. these $x',x''$-components provide us exactly with the necessary semiclassical corrections and show simultaneously the physical meaning of the latter.

Therefore the energy levels \mref{121} reproduce the ones in \mref{116a} by
\be
E_{ck+nr,cl-nq}=\frac{\pi^2}{2}\ll(\frac{m^2}{q^2a^2+r^2b^2}+\frac{n^2r^2}{a^2\sin^2\alpha_{qr}}\r)
\label{124}
\ee
where $m=c(qk+rl)$.

Consider now the SWF \mref{119} for the quantum numbers \mref{123} corresponding to the periodic skeletons on Fig.4 we have
\be
\Psi_{m,n}^{(per)}(x,y)\equiv\Psi_{ck+nr,cl-nq}^{(sem)}(x,y)=-4A'\sin(p_x^{(per)}x)\sin(p_y^{(per)}y)=\nn\\
A'\ll(e^{i{\bf p}^{(per)}{\bf r}}+e^{-i{\bf p}^{(per)}{\bf r}}-e^{i{\bf p}^{(per)}{\bf r}'}-e^{-i{\bf p}^{(per)}{\bf r}'}\r)=\nn\\
A'\ll(e^{i\frac{2\pi m}{D_{qr}}y'}e^{i\frac{\pi nr}{a\sin\alpha_{qr}}x'}+
e^{-i\frac{2\pi m}{D_{qr}}y'}e^{-i\frac{\pi nr}{a\sin\alpha_{qr}}x'}-\r.\nn\\
e^{i\frac{2\pi m}{D_{qr}}y''}e^{-i\frac{\pi nr}{a\sin\alpha_{qr}}x''}-
\ll.e^{-i\frac{2\pi m}{D_{qr}}y''}e^{i\frac{\pi nr}{a\sin\alpha_{qr}}x''}\r)=\nn\\
-2A'\ll(\Psi_{m,n}^{B-S;D}(x,y)-\Psi_{m,n}^{B-S;N}(x,y)\r)
\label{126}
\ee
where
\be
\Psi_{m,n}^{B-S;D}(x,y)=\sin\ll(\frac{2\pi m}{D_{qr}}y'\r)\sin\ll(\frac{\pi nr}{a\sin\alpha_{qr}}x'\r)+
\sin\ll(\frac{2\pi m}{D_{qr}}y''\r)\sin\ll(\frac{\pi nr}{a\sin\alpha_{qr}}x''\r)\nn\\
\Psi_{m,n}^{B-S;N}(x,y)=\cos\ll(\frac{2\pi m}{D_{qr}}y'\r)\cos\ll(\frac{\pi nr}{a\sin\alpha_{qr}}x'\r)-
\cos\ll(\frac{2\pi m}{D_{qr}}y''\r)\cos\ll(\frac{\pi nr}{a\sin\alpha_{qr}}x''\r)
\label{127a}
\ee

\subsubsection{The Bogomolny-Schmit superscar states in the rectangle}

\hskip+2em The SWF $\Psi_{m,n}^{(per)}(x,y)$ in the representation \mref{126} has the following clear structure
\begin{enumerate}
\item it is the coherent sum of the Bogomolny-Schmit superscar states contributing to \linebreak $\Psi_{m,n}^{B-S;D}(x,y)$ and coming from all the POCs having the
period ${\bf D}_{qr}$ as well as from the POCs with the period ${\bf D}_{-qr}$ but only in the half;
\item it is completed by another contribution represented by $\Psi_{m,n}^{B-S;N}(x,y)$ which is also the coherent sum of states defined in a way similar
to the Bogomolny-Schmit superscar ones on both the kind of POCs mentioned in the previous point but satisfying on boundaries of these POCs
the Neumann conditions.
\end{enumerate}

It is therefore clear that none of the Bogomolny-Schmit superscar states contributing to $\Psi_{m,n}^{B-S;D}(x,y)$ nor any of their linear combinations
can provide us with a well defined
stationary quantum state in the rectangle. Such a state can be constructed only by a coherent sum of such states contributing to $\Psi_{m,n}^{B-S;D}(x,y)$
and satisfying the Dirichlet conditions on the POC boundaries and the superscar states contributing to $\Psi_{m,n}^{B-S;N}(x,y)$ and satisfying the
Neumann boundary conditions.

Moreover the superscar states of Bogomolny and Schmit contribute only to the states defined on the periodic skeleton which do not exhaust the whole
spectrum of the semiclassical states in the rectangular billiards defined by \mref{119} and therefore they cannot pretend to substitute the full set of
the rectangular SWFs as it was suggested by their inventors \cite{46}.

Further the representation \mref{126} clarifies why the periodic properties of these states which are different from the bouncing ball ones are
difficult to be observed in the real pattern of the SWFs \mref{126} in the rectangle. Of course this is because $\Psi_{m,n}^{(per)}(x,y)$ in such cases
are the sum of the two terms $\Psi_{m,n}^{B-S;D}(x,y)$ and $\Psi_{m,n}^{B-S;N}(x,y)$ which satisfy on the singular diagonals of the respective POCs the
opposite boundaries conditions by which each of the terms destroys the visible boundary effects of the other. The bouncing ball states \mref{23a} are the
only ones which are deprived of the term $\Psi_{m,n}^{B-S;N}(x,y)$ just allowing the Bogomolny-Schmit superscar state $\Psi_{m,n}^{B-S;D}(x,y)$ to be
fully exposed in these cases giving the well known patterns of the standing waves \mref{23} in the rectangular billiards.

\subsection{The broken (L-shape) rectangular billiards}

\hskip+2em This kind of RPB is particularly interesting because of experiments done with the microwave cavities of this form \cite{47}-\cite{51}. Below we will analyse such
a billiard building in it a SWF on periodic skeleton shown in Fig.5.

\begin{figure}
\begin{center}
\psfig{figure=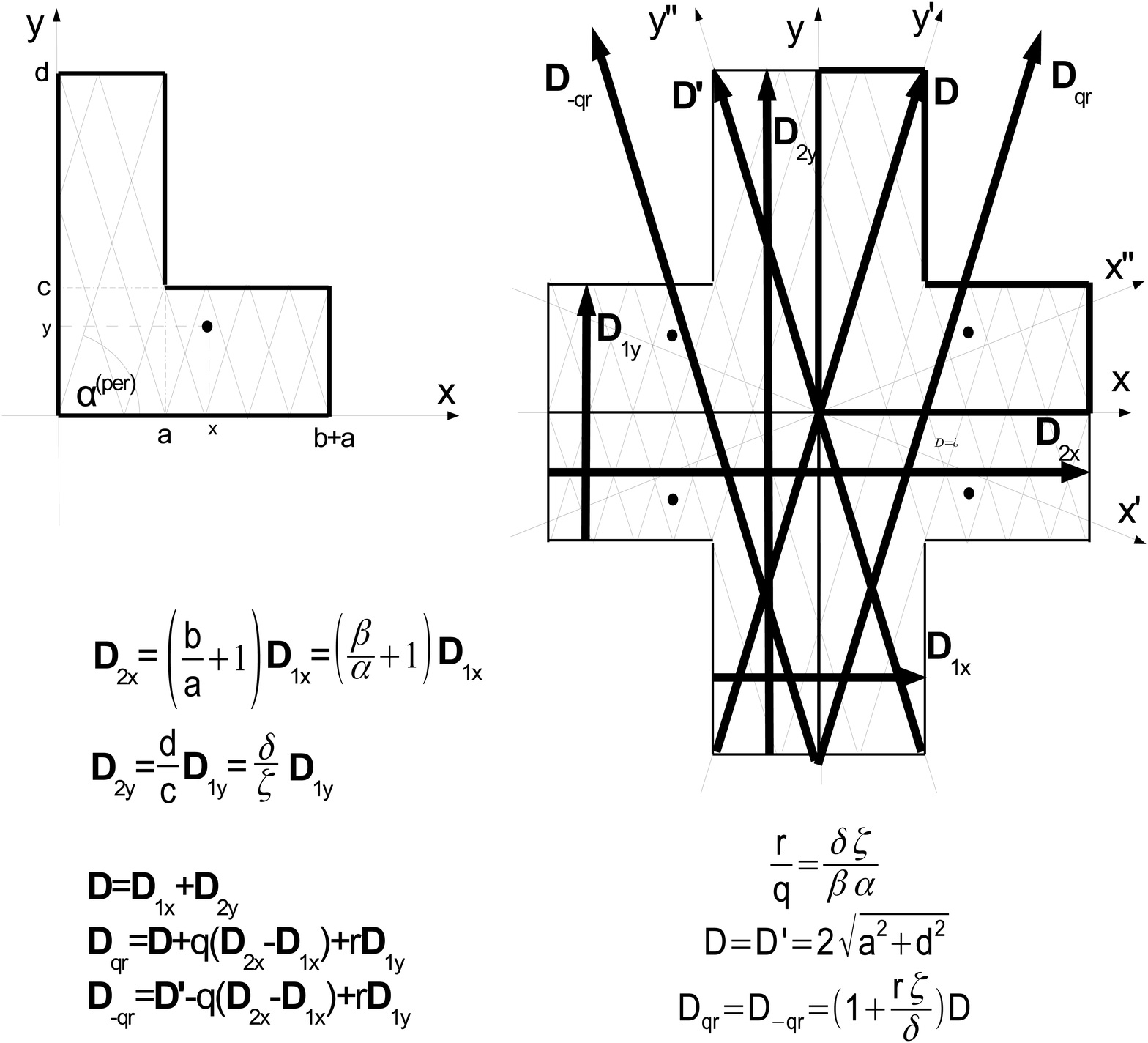,width=10 cm} \caption{The L-shape billiards (left) and its EPP (right). The periods ${\bf D},\;{\bf D}'$ and
${\bf D}_{qr},\;{\bf D}_{-qr}$ defining respective POCs are shown together with singular diagonals as their boundaries}
\end{center}
\end{figure}

First however let us write a SWF built on an aperiodic skeleton with the momentum {\bf p}. The L-shape billiards considered is a doubly rational one
if its sizes satisfy the following conditions
\be
\frac{b}{a}=\frac{\alpha}{\beta},\;\;\;\;\frac{d}{c}=\frac{\delta}{\zeta}
\label{128}
\ee
where $\alpha,\;\beta$ are coprime integers as well as $\delta,\;\zeta$.

\subsubsection{Quantization on aperiodic skeleton}

\hskip+2em Since the shortest period in the $x$-axis direction is ${\bf D}_{1x}/\alpha$ while in the $y$-direction - ${\bf D}_{1y}/\zeta$ then by the routine
procedure applied earlier we have
\be
{\bf p}\cdot{\bf D}_{1x}=2\pi m\alpha\nn\\
{\bf p}\cdot{\bf D}_{1y}=2\pi n\zeta\nn\\
m,n=0,\pm 1,\pm 2, ...
\label{129}
\ee
so that for the energy spectrum we have
\be
E_{mn}=\frac{\pi^2}{2}\ll(\frac{m^2\alpha^2}{a^2}+\frac{n^2\zeta^2}{c^2}\r)\nn\\
m,n=0,\pm 1,\pm 2, ...
\label{130}
\ee
and for the SWF satisfying the Dirichlet boundary conditions we get (up to a normalization constant)
\be
\Psi_{mn}^{(sem)}(x,y)=\sin\ll(\frac{\pi m\alpha}{a}x\r)\sin\ll(\frac{\pi n\zeta}{c}y\r)
\label{131}
\ee
so that we can limit the quantum numbers $m,n$ to the positive values only.

\subsubsection{Quantization on periodic skeletons}

\hskip+2em Representing the SWF \mref{131} by
\be
\Psi_{mn}^{(sem)}(x,y)=\frac{1}{2i}\ll(e^{i\frac{\pi n\zeta}{c}y}\sin\ll(\frac{\pi m\alpha}{a}x\r)-
e^{-i\frac{\pi n\zeta}{c}y}\sin\ll(\frac{\pi m\alpha}{a}x\r)\r)
\label{132}
\ee
or by
\be
\Psi_{mn}^{(sem)}(x,y)=\frac{1}{2i}\ll(e^{i\frac{\pi m\alpha}{a}x}\sin\ll(\frac{\pi n\zeta}{c}y\r)-
e^{-i\frac{\pi m\alpha}{a}x}\sin\ll(\frac{\pi n\zeta}{c}y\r)\r)
\label{133}
\ee
we immediately recognize in the above forms the quantizations on the periodic skeleton identified typically as the bouncing ball one with trajectories parallel
to the $y$-axis in the first of the above cases or on the bouncing ball skeleton with trajectories parallel to the $x$-axis in the second one.

Both the latter quantizations give the same results as for the aperiodic case since none condition on the L-shape billiards sizes is necessary for these
quantizations.

The latter property of the bouncing ball skeletons is in contrast with the quantization on the POCs defined by the periods ${\bf D}$ and ${\bf D}'$.
Trajectories which do not belong to the
latter POCs but are parallel to the periods ${\bf D}$ and ${\bf D}'$ also form POCs, i.e. they are periodic with the respective periods
${\bf D}_{qr}={\bf D}+q({\bf D}_{2x}-{\bf D}_{1x})+r{\bf D}_{1y}$ and  ${\bf D}_{-qr}={\bf D}'-q({\bf D}_{2x}-{\bf D}_{1x})+r{\bf D}_{1y}$ if the
conditions \mref{128} are satisfied. The coprime integers $q,r$ are defined by
\be
\frac{r}{q}=\frac{\beta\delta}{\alpha\zeta}
\label{134}
\ee

If the momentum ${\bf p}^{(per)}$ is directed along the period ${\bf D}$ the respective quantization rules for the momentum on the periodic skeletons
defined by the periods ${\bf D}$ and ${\bf D}_{qr}$ are
\be
{\bf p}^{(per)}\cdot{\bf D}_{1x}= p_x^{(per)}2a=2\pi m'\alpha\nn\\
{\bf p}^{(per)}\cdot{\bf D}_{1y}= p_y^{(per)}2c=2\pi n'\zeta\nn\\
{\bf p}^{(per)}\cdot{\bf D}=2p\sqrt{a^2+d^2}={\bf p}^{(per)}\cdot({\bf D}_{1x}+{\bf D}_{2y})=2\pi (m'\alpha+n'\delta)\equiv 2\pi m''\nn\\
m',n'=0,\pm 1,\pm 2, ...
\label{135}
\ee

However the above quantization is possible under the following condition for the sizes $a,c$
\be
\frac{c^2}{a^2}=\frac{k}{l}\frac{\zeta^2}{\alpha\delta}
\label{136}
\ee
where $k,l$ are coprime integers and $m'=\gamma k,\;n'=\gamma l,\;\gamma=0,\pm 1,\pm 2,...$ .

The respective semiclassical correction $E_0$ to the energy levels is determined by the conditions
\be
\sqrt{2E_0}\frac{D_{1x}}{\alpha}\sin\alpha^{(per)}=2\pi n''\nn\\
\sqrt{2E_0}\frac{D_{1y}}{\zeta}\cos\alpha^{(per)}=2\pi n_1\nn\\
\label{137}
\ee
which can be satisfied if $n''=\omega\delta$ and $n_1=\omega\alpha,\;\omega=0,\pm 1,\pm 2, ...$ .

Therefore the energy spectrum related to the states defined on the periodic skeleton considered is
\be
E_{m''n''}^{(per)}=\frac{\pi^2}{2}\ll(\frac{m''^2}{a^2+d^2}+\frac{n''^2\alpha^2}{\sin^2\alpha^{(per)}}\r)
\label{138}
\ee
while the corresponding BSWF is
\be
\Psi^{BSWF}(x',y')=e^{i\frac{\pi m''}{\sqrt{a^2+d^2}}y'}\ll(A\sin\ll(\frac{\pi n''\alpha}{\sin\alpha^{(per)}}x'\r)+
B\cos\ll(\frac{\pi n''\alpha}{\sin\alpha^{(per)}}x'\r)\r)
\label{139}
\ee

It should be clear that the energy spectrum of the states defined on the skeleton determined by the periods  ${\bf D}'$ and ${\bf D}_{-qr}$ coincides with
\mref{138} while the corresponding BSWF is
\be
\Psi^{BSWF}(x'',y'')=e^{i\frac{\pi m''}{\sqrt{a^2+d^2}}y''}\ll(A\sin\ll(\frac{\pi n''\alpha}{\sin\alpha^{(per)}}x''\r)+
B\cos\ll(\frac{\pi n''\alpha}{\sin\alpha^{(per)}}x''\r)\r)
\label{140}
\ee

Now we can repeat the tricks of the previous section projecting the momentum {\bf p} on the axes $x',y'$ and $x'',y''$ and putting in \mref{130}-\mref{131}
\be
m=\gamma k+n''\delta\nn\\
n=\gamma l-n''\alpha\nn\\
m''=m\alpha+n\delta
\label{141}
\ee
by which we recover the energy spectrum \mref{138} and we get the following representation of the SWF built on the periodic skeleton determined by the
periods  ${\bf D}$ and ${\bf D}_{qr}$
\be
\Psi_{m''n''}^{(per)}(x,y)=\Psi_{\gamma k+n''\delta,\gamma l-n''\alpha}^{(sem)}(x,y)=\fr(\Psi_{m'',n''}^{B-S;D}(x,y)-\Psi_{m'',n''}^{B-S;N}(x,y))
\label{142}
\ee
where
\be
\Psi_{m'',n''}^{B-S;D}(x,y)=\sin\ll(\frac{2\pi m''}{D}y'\r)\sin\ll(\frac{\pi n''\alpha}{a\sin\alpha^{(per)}}x'\r)+
\sin\ll(\frac{2\pi m''}{D}y''\r)\sin\ll(\frac{\pi n''\alpha}{a\sin\alpha^{(per)}}x''\r)\nn\\
\Psi_{m'',n''}^{B-S;N}(x,y)=\cos\ll(\frac{2\pi m''}{D}y'\r)\cos\ll(\frac{\pi n''\alpha}{a\sin\alpha^{(per)}}x'\r)-
\cos\ll(\frac{2\pi m''}{D}y''\r)\cos\ll(\frac{\pi n''\alpha}{a\sin\alpha^{(per)}}x''\r)
\label{143}
\ee

The form \mref{142} of $\Psi_{m''n''}^{(per)}(x,y)$ shows that all the final statements made in the case of the rectangular billiards can be repeated
also in the case of the L-shape billiards considered. In particular the valid one is the statement that only the bouncing ball patterns can be clearly
visible in the respective experimental observations of the stationary states in the L-shape billiards done by Sridhar \cite{47} Kudrolli and
Sridhar \cite{48} and Sridhar and Heller \cite{51}.

\section{Summary and conclusions}

\hskip+2em In this paper we have constructed the semiclassical wave functions in the rational polygon billiards by the method of Maslov and Fedoriuk
\cite{4} adapted just to these cases of the quantum systems \cite{43}-\cite{42}.

The SWFs constructed for the billiards having the particular
forms of the right triangle, the parallelogram, the rectangle and the L-shape have been analysed in the context of their relations with the state
considered by Bogomolny and Schmit \cite{46} and known as the superscar waves propagated in POCs.

We have established the following rather general facts
\begin{enumerate}
\item SWFs satisfying standard conditions of the quantum mechanics can always be built on the aperiodic skeletons and also but with possible restrictions
on the periodic ones;
\item if the constructions of SWFs on the periodic skeletons are not restricted then these SWFs fully coincide with the ones built on the aperiodic
skeletons - in other cases the respective SWFs reproduces only part of the full set of solutions built on the aperiodic skeletons, i.e. an energy
spectrum of the states built on the periodic skeletons is in general only a subset of the full one;
\item periodic skeletons are built of POCs;
\item SWFs built on the periodic skeletons are always coherent sums of contributions from all of their component POCs;
\item a respective contribution to SWFs built on the periodic skeleton and coming from a particular POC can behave as the Bogomolny - Schmit superscar
states for a subset of energies covered by the SWFs but it looses any contact with such states for other energies;
\item the Bogomolny - Schmit superscar states cannot exist in general as such in RPBs - one can meet them as separate states in some exceptional
cases only such as the states in the rectangular billiards, in the equilateral triangle or in the L-shape billiards;
\item the Bogomolny - Schmit superscar states are not the only ones which can contribute to SWFs built on the periodic skeletons - there are also another
superscar contributions to these SWFs which satisfy on the POC boundaries rather the Neumann conditions than the Dirichlet ones;
\item the superscar phenomena can be observed only for the states which are built on the periodic skeletons and only then which is not a rule however they
can reveal a POC structure of periodic skeletons;
\item the Bogomolny - Schmit superscar states can be considered in general as a useful notion in the description of the SWF structure built on
the periodic skeletons rather than a notion describing real independent quantum states in RPBs.
\end{enumerate}

\appendix

\section{Structure of the period space $V_{2g}^{DRPB}$ projected on the plane in the case of DRPB}

\hskip+2em Assume the situation described in the point 19. in sec.2. Multiplying \mref{1c} by $q_{2j}p_j$ we get
\be
\frac{p_{1j}q_{2j}q_j}{q_{1j}}{\bf D}_1={\bf D}_1'=-q_{2j}p_j{\bf D}_j-p_{2j}q_j{\bf D}_k
\label{A1}
\ee
where ${\bf D}_1'$ is a vector in $V_{2g}^{DRPB}$.

It is now easy to show that the vector
\be
{\bf d}_1=\frac{{\bf D}_1}{q_{1j}}
\label{A2}
\ee
is also a period, i.e. it belongs to $V_{2g}^{DRPB}$.

For this goal let us assume for a convenience that $p_{1j}q_{2j}q_j>q_{1j}$. Then we can write $p_{1j}q_{2j}q_j=a_1q_{1j}+b_1,\;b_1<q_{1j}$, so that
\be
{\bf D}_1''=\frac{b_1}{q_{1j}}{\bf D}_1
\label{A3}
\ee
is again a vector in $V_{2g}^{DRPB}$.

Writing ${\bf D}_1=\frac{q_{1j}}{b_1}{\bf D}_1''$ and putting $q_i=a_2b_1+b_2,\;0<b_2<b_1$, we can repeat the previous procedure. The procedure stops at
the $k$-th step for $k<q_{1j}$ when $b_{k-1}=1$ so that
\be
{\bf D}_1^{(k)}=\frac{1}{b_{k-2}}{\bf D}_1^{(k-1)}
\label{A4}
\ee
is again in $V_{2g}^{DRPB}$.

Therefore we have
\be
{\bf d}_1=\frac{1}{q_{1j}}{\bf D}_1=\frac{1}{b_1}{\bf D}_1''=\frac{1}{b_2}{\bf D}_1^{(3)}=...=\frac{1}{b_{k-2}}{\bf D}_1^{(k-1)}={\bf D}_1{(k)}
\label{A5}
\ee
which proves our assertion.

In a similar way we can prove that ${\bf D}_1/C_1$ is also a vector in $V_{2g}^{DRPB}$ if $C_1$ is the least common multiple of
$q_{1j},\;j=3,...,2g-2$.

\end{document}